# Ferroelectric polarization mapping through pseudosymmetry-sensitive EBSD reindexing

Claire Griesbach[1], Tizian Scharsach[2], Morgan Trassin[2], Dennis M. Kochmann[1]

[1] Mechanics & Materials Laboratory, Department of Mechanical and Process Engineering, ETH Zürich, 8092, Zürich, Switzerland

[2] NEAT Lab, Department of Materials, ETH Zürich, 8093, Zürich, Switzerland

**Abstract**

Ferroelectric materials exhibit a switchable, spontaneous polarization at the unit cell level—an attractive property utilized in many emerging technologies for, among others, high-density memory storage, low-power transistors, and high-speed fiber optic communication. Understanding the local polarization switching behavior, through domain nucleation and evolution, is critical to advancing these technologies and requires characterization of the local domain microstructure. However, in application-relevant polycrystalline materials exhibiting a distribution of grain orientations, a direct mapping of the polarization direction in three dimensions has remained inaccessible using conventional experimental approaches. Here, taking barium titanate single crystals and lead zirconium titanate polycrystals as our bulk model systems, we map the local polarization directions using a new electron backscatter diffraction indexing technique based on simulated pattern-matching. Through *improved pre-processing techniques* (including optimized pattern processing, a new pseudosymmetry-sensitive neighbor pattern averaging method, and DIC-based global sample-detector geometry calibration) and a *new pseudosymmetry confidence index* (which considers not only pattern similarity but pattern dissimilarity trends with other domain variant patterns), we successfully distinguish between the six polarization directions, despite the challengingly small unit cell aspect ratio of the selected materials. The methods developed in this work are not only applicable to ferroelectrics but any material which exhibits close crystallographic pseudosymmetries—extending the current capabilities of EBSD.

# 1. Introduction

Ferroelectrics are active materials that exhibit spontaneous polarization within the crystal unit cell, which can be reoriented by an imposed electric field or stress. This property, in addition to their piezoelectric and pyroelectric properties, makes ferroelectric materials a versatile and powerful tool for applications including radio and microwave devices [1], sensors and actuators [2,3], infrared imaging [4], waste-heat energy conversion [5], next-generation transistors [6,7], high-density data storage devices [8], solar cells [9], and high-speed fiber optic communication [10]. Many of these applications rely on bulk ferroelectric properties which are dictated by microstructural evolution. In ferroelectrics, the microstructure contains regions of crystallographic unit cells with the same polarization direction (called domains), which are separated by domain walls. Bulk polarization reversal is accommodated through local switching of the domain microstructure through the nucleation of new domains and motion of domain walls. Understanding the nucleation and evolution of domains due to applied electric fields or stress is critical to advancing ferroelectric technologies.

Several nondestructive methods—both *in situ* and *ex situ*—have been used to characterize ferroelectric domain microstructures, including piezoresponse force microscopy (PFM) [11–14], optical methods [15], electron microscopy [16], and x-ray and neutron diffraction [17–20]. Optical methods are typically limited in spatial resolution, and often the domain orientations cannot be ascertained [15,21]. However, optical methods are easiest to integrate with experimental setups to observe domain evolution in real time [22–24]. In-situ x-ray and neutron diffraction experiments typically only provide the average crystallographic or bulk switching response [17–19]. In contrast, fine domain structures can be spatially resolved through PFM and electron microscopy techniques due to their nanometer spatial resolutions.

PFM is among the most widely used techniques for imaging ferroelectric domain microstructures. It utilizes the converse piezoelectric effect to create a map of domains based on the varying magnitudes of surface displacement upon an applied electric field. When several PFM scans are taken at different sample orientations, the different lateral and vertical PFM signals can be deconvoluted and combined to visualize the polarization directions [11]. This technique, referred to as vector-PFM, poses several challenges for high-quality data collection—including macroscopically rotating the sample while finding the same microscopic area, obtaining calibration coefficients, and signal noise [14]. Additionally, without knowledge of the crystal orientation, it is difficult to accurately calibrate the intensity in the different signals to the polarization direction. Therefore, PFM is typically used on single crystals where the crystal orientation is known or on polycrystals as a non-quantitative visualization technique.

For most crystalline materials, obtaining spatially-resolved crystal orientations is accomplished through electron microscopy (EM) techniques. In the scanning electron microscope (SEM), electron backscatter diffraction (EBSD) provides maps of the local crystal orientations at spatial resolutions down to 10s of nanometers. However, EBSD and EM in general are underutilized in ferroelectrics. There are two main challenges which have limited the usage of EM in ferroelectrics. The first is that the electron beam can cause domain evolution if charging occurs [25,26], making it difficult to characterize a stable domain microstructure. The second reason pertains specifically to using EBSD to identify domain orientations: depending on the *c*/*a*-ratio of the crystallographic unit cell, there may be only minute differences between Kikuchi diffraction patterns of polar domains (polarization directions rotated by 180°). These differences are from dynamic diffraction (multiple scattering events), not from geometrically-defined Bragg diffraction—meaning typical Hough-based EBSD indexing techniques cannot identify domain orientations [27,28]. If EBSD is included in a study of ferroelectrics, it is usually used only to determine the grain orientations using traditional Hough-based indexing or to identify 90° domain laminates [28–30]. There are newer EBSD indexing methods that use dynamically simulated diffraction patterns to find the correct orientation—broadly referred to as pattern matching techniques. The most common ones are dictionary indexing and spherical indexing [31–34]. These techniques are theoretically capable of distinguishing between polar domains, since the intensity differences are also captured in the simulated patterns, as recently demonstrated for single-crystal lithium niobate [27]. Unfortunately, other ferroelectrics like lead zirconium titanate (PZT) and barium titanate (BTO) have greater similarities between Kikuchi patterns of different domain variants, so that pattern matching fails. Furthermore, using pattern matching indexing to identify domains in polycrystalline ferroelectric materials has not been accomplished to date. In this study, we develop a new EBSD reindexing approach, which successfully identifies the local polarization direction in ferroelectric domain microstructures. This approach is not only applicable to ferroelectric materials but any material which exhibits pseudosymmetry (PS) challenges in EBSD indexing.

This work focuses on two challenging ferroelectrics for EBSD investigation, BTO and PZT. However, the developed methods are broadly applicable to pseudosymmetric materials. In Section 2, we demonstrate the limitations of the state-of-the-art pattern pre-processing and indexing methods and the challenges surrounding EBSD of ferroelectrics. In that section, we also present an automatic pattern processing optimization routine, a PS-sensitive neighbor pattern averaging algorithm, and a DIC-based global geometry refinement algorithm. All these improved methodologies are particularly tailored to EBSD of PS materials, including ferroelectrics. In Section 3, we present our new reindexing technique, which takes into account not only pattern similarity but also pattern dissimilarity trends with other PS variants to significantly improve the discernability between variants. In Section 4, we demonstrate successful indexing of a single-crystal BTO sample, validated with PFM, and a polycrystalline PZT sample. This work enables accurate determination of the spatially varying polarization directions and crystal orientations in ferroelectrics—key microstructural information currently unattainable with other commonly used techniques. Furthermore, this work extends the capabilities of EBSD to accurately distinguish between close PS variants.

## 2. Overcoming challenges of EBSD on ferroelectrics

In this section, we establish the challenges and limitations of the state-of-the-art EBSD data processing and indexing methods, as applied to ferroelectrics. We propose several new data acquisition and processing methods, which are essential for EBSD analysis of ferroelectrics and other pseudosymmetric materials.

### 2.1. Beam sensitivity: from inadvertent domain evolution to harnessing domain contrast

A primary challenge of using EM to study ferroelectrics is the innate sensitivity to the electron beam. As with all insulating materials in the SEM, lack of a conductive pathway to ground can cause charge build-up on the sample surface and a local electric field because of the voltage differential between the top and bottom of the sample. Since ferroelectrics have the intrinsic property of switchable polarization, this electric charging can cause the local domain structure to evolve. Thus, it is critical to mitigate charging as much as possible to avoid domain evolution. Figure 1A shows examples of SEM images taken of the same few grains before (A-1) and after (A-2) the region was exposed to charging, showing significant evolution of the domain structures within several grains. (For both SEM imaging and EBSD data acquisition, we use a TFS Quanta 200F SEM with the chamber vacuum pressure at 30 – 60 Pa.) To avoid charging-induced domain evolution, one would ideally operate under charge-neutral conditions, where the emitted electrons equal the incident electrons (Figure 1(A-3)). Realistically, achieving this perfect charge-neutral condition by only adjusting the accelerating voltage is difficult, which is why other measures are typically taken such as adding a conductive coating to the sample or operating in a low-vacuum SEM chamber.

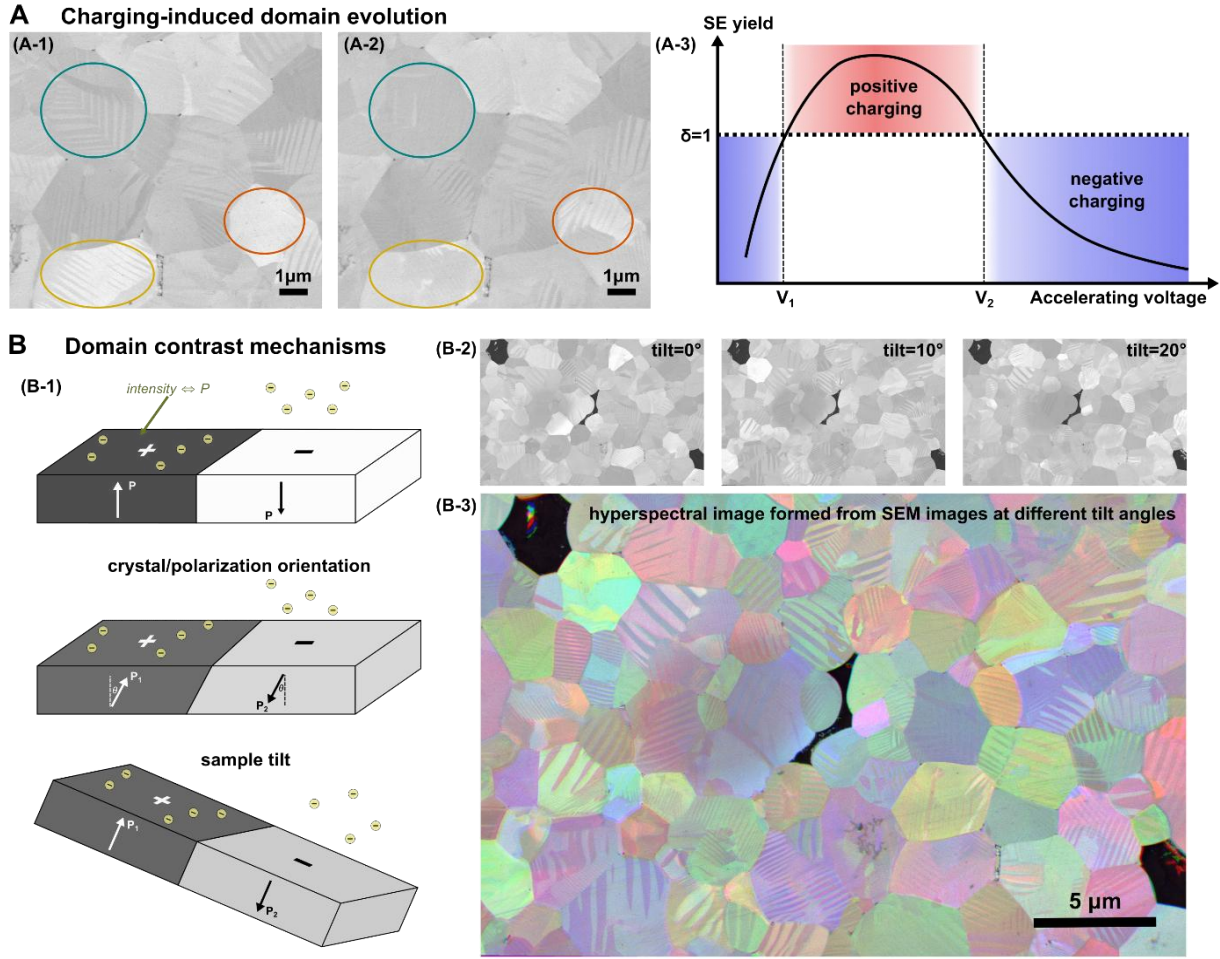

**Figure 1: Scanning electron microscopy of ferroelectrics:** (A) Charging-induced domain evolution in a PZT polycrystal: domain evolution observed between SEM images (A-1) and (A-2), (A-3) schematic adapted from [16], showing positive, negative, and charge-neutral conditions, considering secondary electron yield (emitted/incident) as a function of accelerating voltage; (B) domain contrast mechanisms shown in (B-1) schematics, (B-2) SEM images showing domain contrast changes as a function of tilt angle, and (B-3) hyperspectral image revealing full domain microstructure of a polycrystalline PZT sample.

To mitigate charging further during EBSD data acquisition, the sample is coated with 2 nm of carbon. However, to locate the area with the domain microstructure of interest, SEM imaging was performed on the sample prior to coating with carbon, since a conductive coating obscures the domain contrast. The conductive material has a significantly higher electron yield than the underlying ferroelectric material, obscuring the small electron yield differences between different domains. A potential domain contrast mechanism, as explained in [16] and illustrated in Figure 1(B-1), arises from the bound surface charges associated with the local ferroelectric polarization. In a 180° domain pair, one domain has a polarization pointing upward, producing positive bound charge at the surface, while the adjacent domain has a downward polarization and negative surface charge. These differences in surface charge modify the escape probability and detection of secondary electrons: regions with positive surface charge tend to attract low-energy emitted electrons and reduce the detected electrons (lower intensity), whereas regions with negative charge repel more electrons toward the detector (higher intensity). As a result, domains of opposite polarization can exhibit differing secondary-electron intensities in the SEM image. If the local polarization orientations change, as in the second panel of Figure 1(B-1), the relative intensity also changes. Tilting the sample is analogous to rotating the crystal, also causing intensity changes. Under some orientations, the normal component of the polarization directions between neighboring domains may be nearly identical and no contrast is seen between the two domains.

We have used this knowledge of the domain contrast mechanism to establish a *new qualitative visualization technique* for ferroelectric microstructures. SEM images are taken at various sample tilts or rotations to reveal different domains. (Figure 1(B-2) shows examples of SEM images taken at three different tilt angles of a PZT ferroelectric ceramic.) The images are tilt-corrected, aligned, and cropped to one area. Each image is then considered to be a different channel of information within a hyperspectral dataset. A hypercube object is generated in *Matlab* and used to colorize the data into one RGB image, as shown in Figure 1(B-3). This method can be used to qualitatively visualize and analyze domain microstructures.

As demonstrated, it is essential to tailor the SEM parameters carefully for imaging and EBSD analysis of ferroelectric materials. In this work, we have studied BTO single crystals and PZT polycrystals. As the domain structures are smaller in the studied PZT polycrystals, it is easier to switch them with local electric field build-up from charging, so special care must be given to avoid charging as much as possible. For the bulk BTO single crystal, the domain structure is more stable, so the accelerating voltage can be increased to optimize domain contrast and Kikuchi pattern quality. The parameters typically used for imaging and EBSD analysis on both materials are given in the Supplementary Information, Section 1. These parameters can also be used as a starting point for other ferroelectric materials and SEMs, but care should be taken to test and adjust them until no domain evolution is observed.

*2.2. Kikuchi diffraction patterns of ferroelectric domain variants and pattern matching techniques*

Recent advancements in simulating Kikuchi diffraction patterns have yielded a variety of powerful pattern matching techniques, which can be broadly placed into two categories: spherical indexing [32,33] and dictionary indexing [31,34]. These methods vary in the way the indexing is performed, but both rely on the simulation of a spherical master pattern according to the sample's crystallography and electron beam conditions. These simulations utilize Monte Carlo methods to simulate the stochastic generation of back scattered electrons (BSEs) and Bloch wave analysis to track the dynamic scattering of BSEs within the crystal lattice [35]. Because dynamic scattering events are simulated, the rich intensity information within the bands is visible in the patterns—not just the band geometry and intersections as defined by Bragg diffraction. This information is essential to distinguish between patterns with close pseudosymmetries. In the case of ferroelectric materials, the unit cell is non-centrosymmetric due to the positioning of ion(s), which produces the spontaneous polarization. Flipping this non-centrosymmetric unit cell along the *c*-direction (corresponding to a 180° reversal of the polarization direction) will not alter the band geometry but will show up in intensity differences within the bands. Figure 2A shows two Kikuchi patterns of a BTO crystal, with orientations flipped by 180°. As highlighted by the lines marking the bands, the positioning and intersections of the bands do not change. However, looking closely at the magnified regions of the central pole, the intensities within the bands change. This demonstration shows that it is impossible to distinguish between these two orientations using Hough indexing, which only considers the band geometry.

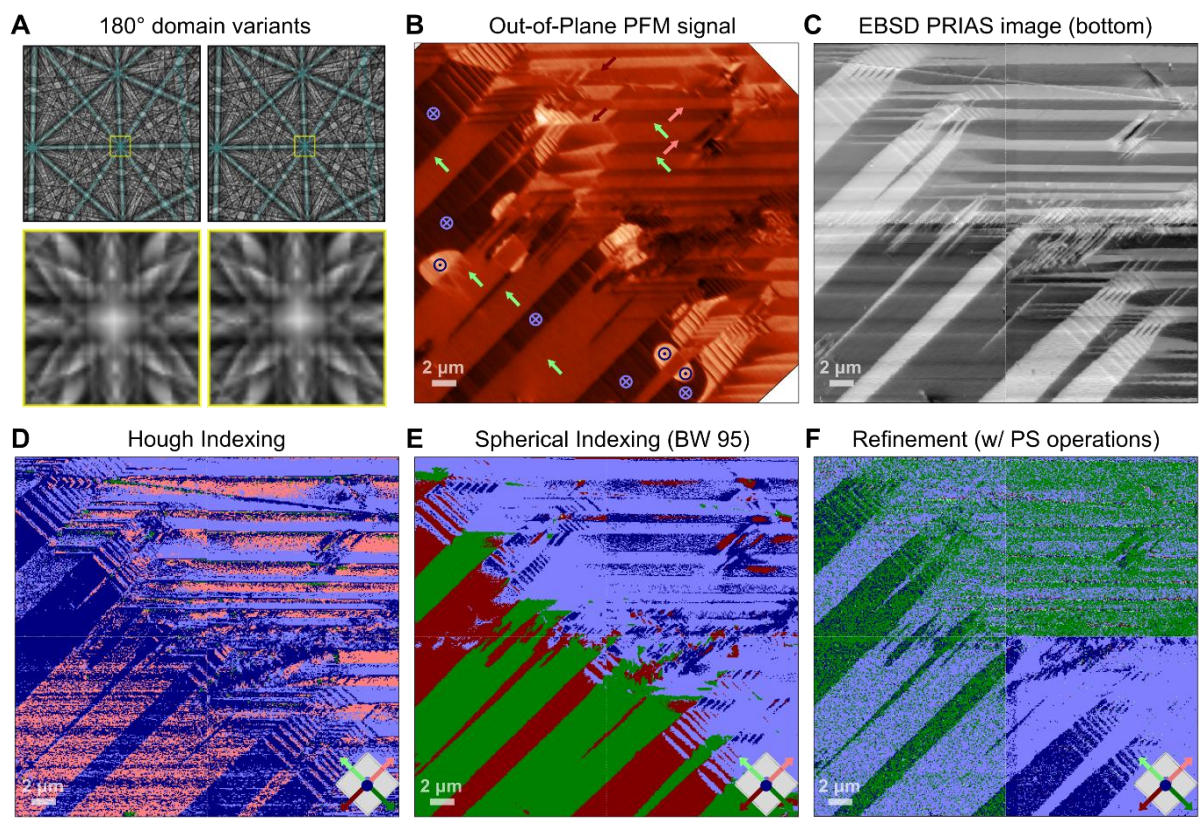

**Figure 2: Current EBSD indexing methods:** (A) Simulated Kikuchi patterns of 180° domain variants of BTO with magnified regions showing intensity differences at the pole; (B) out-of-plane PFM signal of a domain pattern in single-crystal BTO with identified domain orientations shown by arrows (directions identified using in-plane and out-of-plane PFM signals from two different sample orientations; presented in SI, Section 2); (C) Kikuchi pattern **PRIAS** signal showing the same domain microstructure as in the PFM scan; (D) Hough indexing of the EBSD scan; (E) spherical indexing result (using a bandwidth of 95 in OIM9); (F) refinement (in OIM9 with PS operations checked) indexing result. Orientations in D–F are color coded by the domain variant (unit cell color key in the bottom right corner).

To have a chance at determining the polarization orientation using EBSD, we must turn to pattern matching approaches. Developed in the last decade, pattern-matching-based indexing has recently been released in commercial software, including *EDAX's OIM 9* [36] and *Oxford's MapSweeper* [37]. The SEM used for the EBSD analysis presented in this paper is equipped with an *EDAX* camera and associated state-of-the-art *OIM 9* software, which is why we will present the indexing results obtained using two of *OIM 9*'s pattern-matching-based indexing modes for reference: spherical indexing and refinement. Essentially, spherical indexing back-projects each experimental Kikuchi pattern onto the spherical master pattern using a spherical harmonic transform and finds the orientation corresponding to the best match [32]. The precise sample-detector geometry is needed for the back-projection onto the spherical master pattern. Dictionary indexing (and typically refinement) use the inverse of this process: simulated patterns are projected from the spherical master pattern onto a virtual detector (again, the precise sample-detector geometry is required) and the experimental pattern is compared to these simulated patterns to find the best match. Dictionary indexing is significantly more computationally intensive, since each experimental pattern is compared to a large dictionary of simulated patterns. The "refinement" method aims to circumvent this computational bottleneck by beginning with an approximate guess for the orientation (often the Hough indexed orientation) and then searching only within a window of a few degrees in orientation space to find the optimal orientation. To identify the correct orientation, a confidence index (CI) is computed, and the orientation with the maximum confidence index is accepted. For spherical indexing, the CI is the spherical cross correlation as defined in [32]. For dictionary indexing, the CI is typically taken to be the normalized cross correlation (NCC) (although the normalized dot product or unnormalized cross correlation is sometimes used [38,39]):

$$c(\mathbf{A},\mathbf{B}) = \frac{\sum_m \sum_n (A_{mn} - \overline{\mathbf{A}})(B_{mn} - \overline{\mathbf{B}})}{\sqrt{\left[\sum_m \sum_n (A_{mn} - \overline{\mathbf{A}})^2\right]\left[\sum_m \sum_n (B_{mn} - \overline{\mathbf{B}})^2\right]}}, \quad (1)$$

where $A_{mn}$ and $B_{mn}$ are the intensities at pixel $(m,n)$ and $\overline{\mathbf{A}}$ and $\overline{\mathbf{B}}$ are the average intensities for images $\mathbf{A}$ and $\mathbf{B}$. Note that both confidence index definitions are different from that used with Hough indexing, so comparing CI values between the different approaches is meaningless.

We have applied the refinement and spherical indexing methods of *OIM 9* to the EBSD maps of the single-crystal BTO sample analyzed throughout this work. We selected single-crystal BTO as a test case for two reasons: (i) the high pattern similarity between domain variants makes it a challenging material to correctly index via EBSD (see Sections 3.1-3.2), which has not been successfully achieved to date, and (ii) given it is a single crystal, we can identify the correct polarization orientations using PFM to validate our method. We performed PFM and EBSD on the same region of a BTO single crystal showing a complicated domain structure. Figure 2B shows the out-of-plane PFM signal with colored arrows denoting the identified domain orientations. Details of the PFM scanning method, the four signals, and the logic behind the domain identification are presented in the Supplementary Information, Section 2. Next to the PFM scan in Figure 2C is the PRIAS signal map [40] from the EBSD scans, which captures the domain microstructure well. Comparing this image to the PFM signal provides important validation in three ways: (i) the same region is analyzed by PFM and EBSD, (ii) the chosen step size for EBSD is appropriate to resolve the fine domain structures, and (iii) there are sufficient differences between patterns (since the PRIAS signal is calculated from pattern intensities) to distinguish between domains.

Now, we can proceed to analyze the indexing results from the different available methods: Figure 2D shows the Hough indexing solution (obtained during data acquisition), Figure 2E shows the spherical indexing solution (performed at a bandwidth of 95, using a spherical master pattern for BTO and simulated using *EMsoft* [31]), and Figure 2F shows the refinement solution (the 5 PS variants were considered). The variant orientations are colored based on the color keys in the bottom right corners, depicting the crystal orientation through the unit cell and the six possible polarization directions for BTO. In all three indexed maps, the domain microstructure is barely visible due to noise in orientations, and some neighboring domains which show contrast in the PFM and PRIAS images are indexed as the same orientation in the EBSD maps. An exception to this is the region at the bottom left of the spherically indexed map, which shows consistent indexing of the laminate pattern in a dark-red/dark-green alternating pattern. However, the polarization orientations of these domains are not consistent with those identified by PFM (light-blue/light-green pattern). This analysis indicates that the state-of-the-art EBSD indexing techniques are not capable of distinguishing between close PS variants like the ferroelectric domains of BTO. This has motivated the following work to establish a new pattern-matching method with better discrimination between patterns of close pseudosymmetries.

*2.3. Optimization of pattern processing through Bayesian optimization using Gaussian Processes*

Image processing of Kikuchi patterns prior to indexing is common and implemented in all commercial EBSD applications. There are many filters and corrections available, with the most common being a combination of background correction and contrast enhancement. Typically, these are applied in an ad-hoc manner with the user picking an assortment of processing steps and associated parameters based on qualitative examination of a representative pattern. When using pattern matching approaches (where pixel-by-pixel intensities are compared), utilizing the full intensity spectrum and having good feature contrast improves the matching confidence between experimental and simulated patterns. To maximize the matching confidence, we introduce a method to automatically determine the optimal pattern processing parameters.

Our algorithm uses Bayesian optimization with Gaussian processes to select a set of processing parameters which maximizes the normalized cross correlation (NCC) between a matched simulated and experimental pattern. First, a subset of experimental patterns from the map are reindexed using *Kikuchipy's* [41] *refine_orientation_projection_center* function. Patterns are simulated with this new optimized solution, and the NCC between the experimental and simulated patterns is maximized by adjusting the pattern processing parameters. Figure 3A shows a schematic representation of the algorithm. In this work, we chose three pattern processing steps with functions available in *Kikuchipy*: dynamic background subtraction (DBS), adaptive histogram equalization (AHE), and high- and low-pass Fast-Fourier Transform (FFT) filters. DBS was chosen to remove the background intensity variations, AHE was chosen to improve the contrast, the low-pass FFT filter was chosen to reduce noise, and the high-pass FFT filter was chosen to filter out large intensity variations across the detector. Other processing pipelines could be used within the optimization algorithm, with the selection of processing steps depending on the quality of the raw Kikuchi patterns. Given these processing steps, there are eight individual parameters to optimize. For DBS, subtraction of a Gaussian convolution filter in the frequency domain was selected, so two parameters are required—the standard deviation of the Gaussian window and the associated truncation factor. For the AHE, we implemented an on/off switch (an additional parameter) and, if AHE is on, there are three additional associated parameters: the kernel size, clip limit, and number of intensity bins. The FFT filters each have one parameter, which is the cutoff frequency. In total, there is an eight-dimensional parameter space with an undefined functional representation, providing an ideal scenario to apply multivariate Gaussian process approximations. Each of the eight parameters is defined as a search dimension, either as a categorical value or as an integer value within a given range (these are defined explicitly in the Supplementary Information, Section 3). We use the *gp_minimize* function from the *scikit-optimize* python package with 150 evaluations and twelve random restart points to solve for the optimal parameters. The objective to minimize is the negative of the normalized cross correlation ($-c$)—meaning we are solving for the processing parameters that maximize the match between the experimental and simulated patterns.

Figure 3 (B–C) shows the results of the processing parameter optimization applied to an EBSD dataset of a BTO single crystal. Figure 3B shows the per-iteration NCC scores, showing quick convergence to an NCC score of ~0.52 with some noise from random restarts and probing different combinations of parameters. Figure 3C shows the improvement in NCC and pattern quality of an example Kikuchi pattern with each applied processing step utilizing the optimized parameters. For example, the unprocessed pattern has an initial pattern quality of 0.32 (demonstrating poor utilization of the intensity spectrum) and an NCC score of 0.09 when compared to the simulated pattern (orientation and pattern center previously optimized). After all processing steps have been applied with the optimized parameters (last pattern in the row), the image quality improves to 0.95 and the NCC score for the pattern compared to the same simulated pattern improves to 0.52. This analysis demonstrates that even properly matched patterns can have a low NCC score if noise and a background dominate the signal. Optimized pattern processing extracts the actual diffraction signal, improving correlation to the simulated pattern. For materials with close pseudosymmetries, it is essential to maximize the diffraction signal and minimize noise as much as possible.

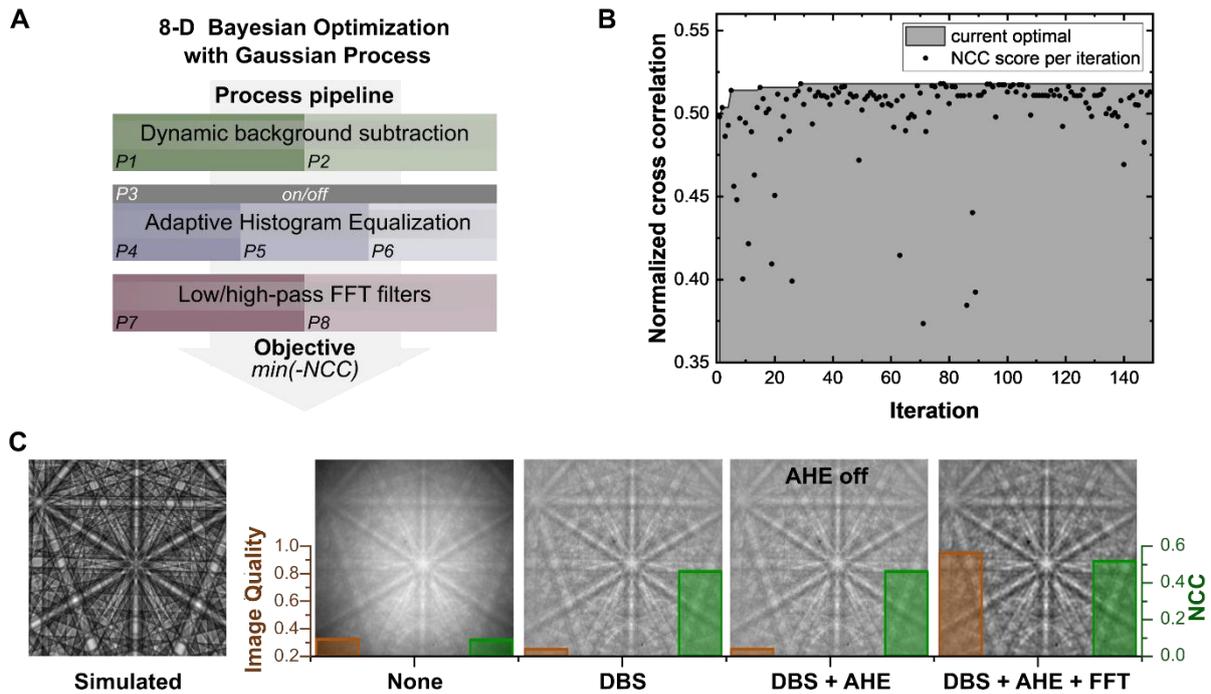

**Figure 3: Optimization of pattern processing:** (A) Schematic showing Bayesian optimization routine for improving the eight pattern processing parameters (P1–P8); (B) optimization of the objective (normalized cross correlation, NCC) per evaluation of the Gaussian process approximation of the function; (C) improvement in image quality and NCC after applying each image processing step with pre-optimized parameters.

*2.4. Pseudo-Symmetry-Sensitive Neighbor Pattern Averaging (PSS-NPA) scheme*

An additional step to reduce noise in experimental patterns is to average neighboring patterns together. This technique was termed Neighbor Pattern Averaging and Re-Indexing (NPAR) by Wright et al. [42]. This approach has been shown to greatly improve pattern quality and indexing success and is implemented in commercial software such as *OIM*. However, a simple NPAR approach causes issues near interfaces of abrupt crystallographic change like grain boundaries or domain walls. To remedy this problem, Brewick et al. [43] introduced the Non-Local Pattern Averaging Reindexing (NLPAR) method, which uses a weighted average based on pattern similarity. A similarity metric, akin to the normalized cross-correlation and accounting for noise, is computed between each pattern and its neighboring patterns within a specified radius. This similarity metric is used as a weight when calculating the averaged pattern, so that only patterns which are similar to the central pattern contribute to the averaging. This method has been shown to work well around grain boundaries, preserving the sharp interface [43]. However, this method begins to break down near interfaces of PS, where the patterns are very similar. Thus, we have developed a new pattern averaging scheme, which we term Pseudo-Symmetry-Sensitive Neighbor Pattern Averaging (PSS-NPA).

The concept behind PSS-NPA is analogous to NLPAR: a weighted average of neighboring patterns is computed using weights determined by pattern similarity. However, the key difference is that we are more selective in which patterns to include. As we show later in Section 3.1, Kikuchi patterns of 180° domain variants in BTO can be up to 99.5% similar. This means that, if using the NCC value as the weight, a weight as high as 0.995 could be applied to a pattern which should be considered different from the central pattern. To solve this, we define a cutoff based on a detectable jump in NCC score—marking the transition from one PS variant to another. First, the NCC is computed for every pattern within the defined radius compared to the central pattern. Then, the NCC scores for all neighbors are sorted in descending order and the moving difference is calculated between each point and the previous point's NCC score ($\Delta c$). To find the first statistically significant jump, we compute statistics of $\Delta c$ within windows before and after each point. If the current $\Delta c$ is larger than the statistical variation within the windows, it is marked as a jump. The first identified jump is taken to be the cutoff: all neighbors before this index are considered in the averaging, while all patterns after are not considered in the averaging. Finally, the central pattern is replaced by an average of itself and the patterns before the cutoff, weighted by their NCC score. An intuitive explanation of this algorithm is presented in Figure 4 (A – B). Figure 4A illustrates a partition of patterns, color-coded by the NCC score with respect to the central pattern. The patterns with a dotted yellow border are those with high NCC scores above the cutoff and are the only neighboring patterns that are to be averaged with the central pattern. The cutoff is determined by the first jump within the sorted NCC values, as shown in Figure 4B. The detailed math behind this algorithm is shown in the Supplementary Information, Section 4. This algorithm has been implemented in *python*, with memory-efficient blocked calculations, which allow for parallel processing.

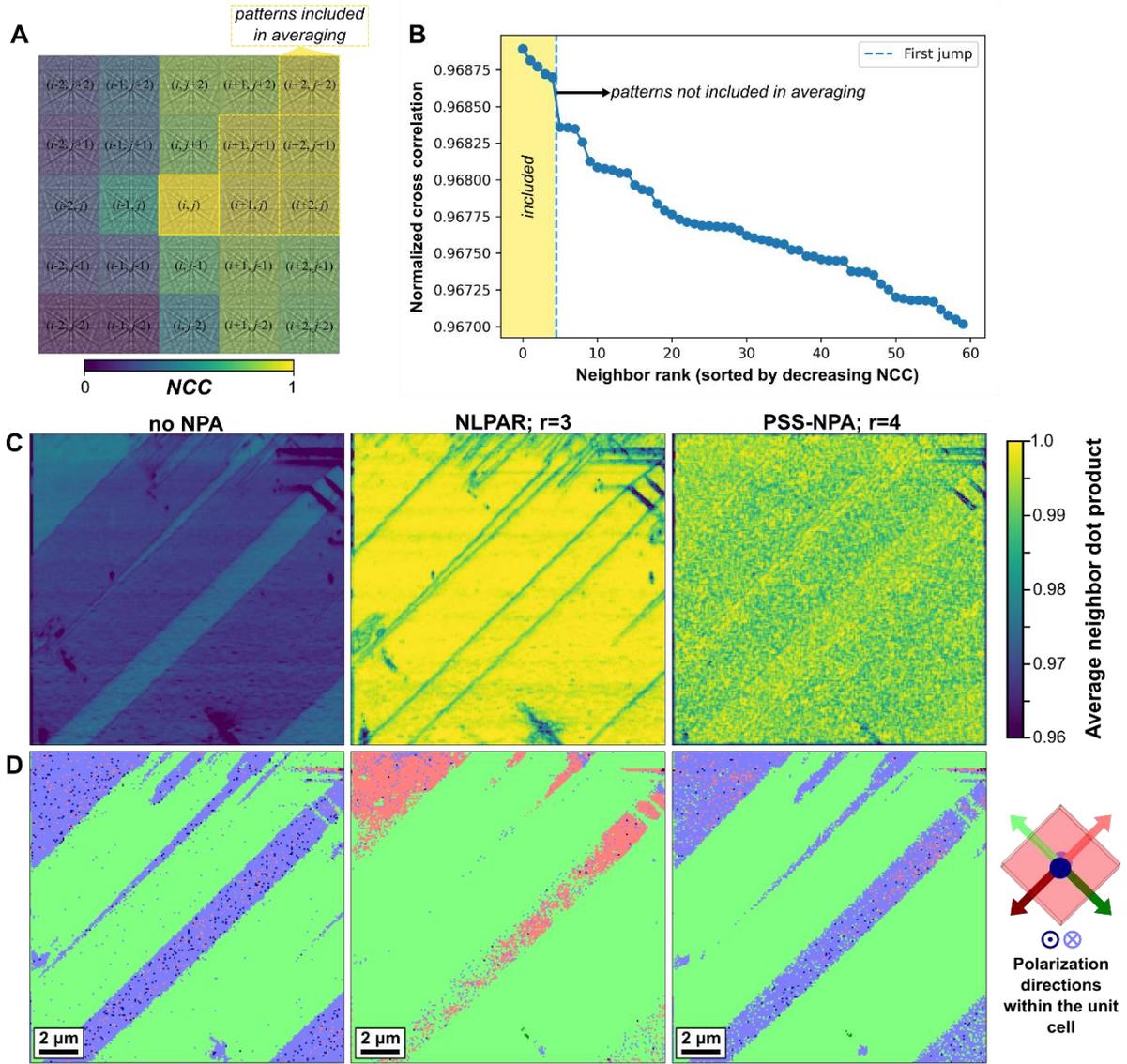

**Figure 4: Pseudo-Symmetry-Sensitive Neighbor Pattern Averaging (PSS-NPA):** (A) Illustration of pattern selection for averaging based on the NCC score with respect to the central pattern and detected PS jump in NCC; (B) plot of NCC vs. neighboring pattern number ordered by NCC, showing a jump in NCC, which is used to cutoff those patterns that are included in the averaging; (C) maps of average neighbor dot product (computed from the four nearest neighbors to each pattern) for pattern sets where no NPA, our PSS-NPA algorithm, and NLPAR were applied; (D) reindexed datasets (orientations shown with domain variant coloring) corresponding to the three NPA methods in (C).

To validate our new PSS-NPA algorithm, we apply it to one of the maps taken on a BTO single crystal (analyzed later in Section 3). Three cases are considered: (i) no neighbor pattern averaging applied, (ii) the NLPAR algorithm applied with a radius of three nearest neighbors, and (iii) our PSS-NPA algorithm applied with a radius of four nearest neighbors. The NLPAR algorithm is applied using the *calcnlpar* function available in the python package *PyEBSDindex* (the value of λ taken to be 1.08 as optimized by the *opt_lambda* function). In all three cases, the other processing and reindexing methods described in this work are applied. To assess the effectiveness of each method, the average neighbor dot product (ANDP: the average of the dot products between each pattern and its four nearest neighbors) maps are plotted in Figure 4C. Both the PSS-NPA and NLPAR methods show a marked increase in ANDP throughout the map, confirming that after averaging neighboring patterns are more similar to one another. However, in comparing the ANDP maps for the two NPA methods, the PSS-NPA method retains more local pattern variation and the boundaries appear sharper than with the NLPAR method. Turning now to the final indexing results (Figure 4D), the domain division is clearly retained with the PSS-NPA method and some noise is eliminated compared to the map where NPA was not applied. However, the reindexed solution with the NLPAR dataset shows disappearance and misidentification of several domains. These results demonstrate that NLPAR is generally not well suited for patterns with close pseudosymmetries. Instead, our new PSS-NPA algorithm retains division between PS variants and improves indexing success.

*2.5. Sample-detector geometry calibration*

The last (but essential) pre-processing challenge that must be overcome is to correctly determine the sample-detector geometry. Precise sample-detector geometry parameters are key to any pattern matching technique, since the projection of simulated patterns from the master pattern relies on this geometry. In fact, any EBSD indexing technique requires knowledge of the sample-detector geometry, with high precision especially required for high-resolution EBSD [44]. Thus, finding the correct parameters is a problem that is as old as the EBSD technique itself. There are six parameters that define the sample-detector geometry (shown in Figure 5A): the sample tilt, detector tilt, detector azimuthal angle, and three parameters characterizing the pattern center location. The pattern center is the point on the detector which is the smallest distance away from the interaction point on the sample. Three values define the pattern center: *pcx* and *pcy* are the *x*- and *y*-coordinates of the pattern center in the detector reference frame, and *pcz* is the distance between sample and detector. The remaining three parameters are angles defining the sample and detector orientations and are typically considered to be fixed and known from the EBSD detector installation or SEM stage positioning readout. Pattern center calibration needs to be performed per EBSD scan or each time the stage position with respect to the detector position is changed. Therefore, pattern center calibration has received much attention, and there are several available methods. Among these are the moving-screen [45], virtual screen [46], shadow-casting [47], single crystal reference [48], projective transformation model [44], integrated digital image correlation [49], homography-based [50], and simulated pattern optimization [51,52] methods. These methods can be grouped into two main categories: (i) "pre-scan calibrants", which are performed system-wide prior to data acquisition, and (ii) "post-scan calibrants", which are performed after data acquisition using the stored Kikuchi patterns. These methods are typically suitable for most EBSD applications; however, for ferroelectrics or other materials with close pseudosymmetries, the accuracy of pre-scan calibration methods—which typically perform the calibration on other reference materials—is insufficient. A new post-scan calibration strategy, called *pcglobal* [51,53], utilizes a global optimization algorithm to simultaneously refine the orientation and pattern center to achieve the best match between experimental and simulated patterns. Convergence to the best pattern center for the map is accomplished by taking the average pattern center over many independently optimized patterns in the map. For the case of PS crystals, refinement is repeated within each PS orientation zone, and the pattern center is selected from the PS variant solution with the maximum correlation to the experimental pattern.

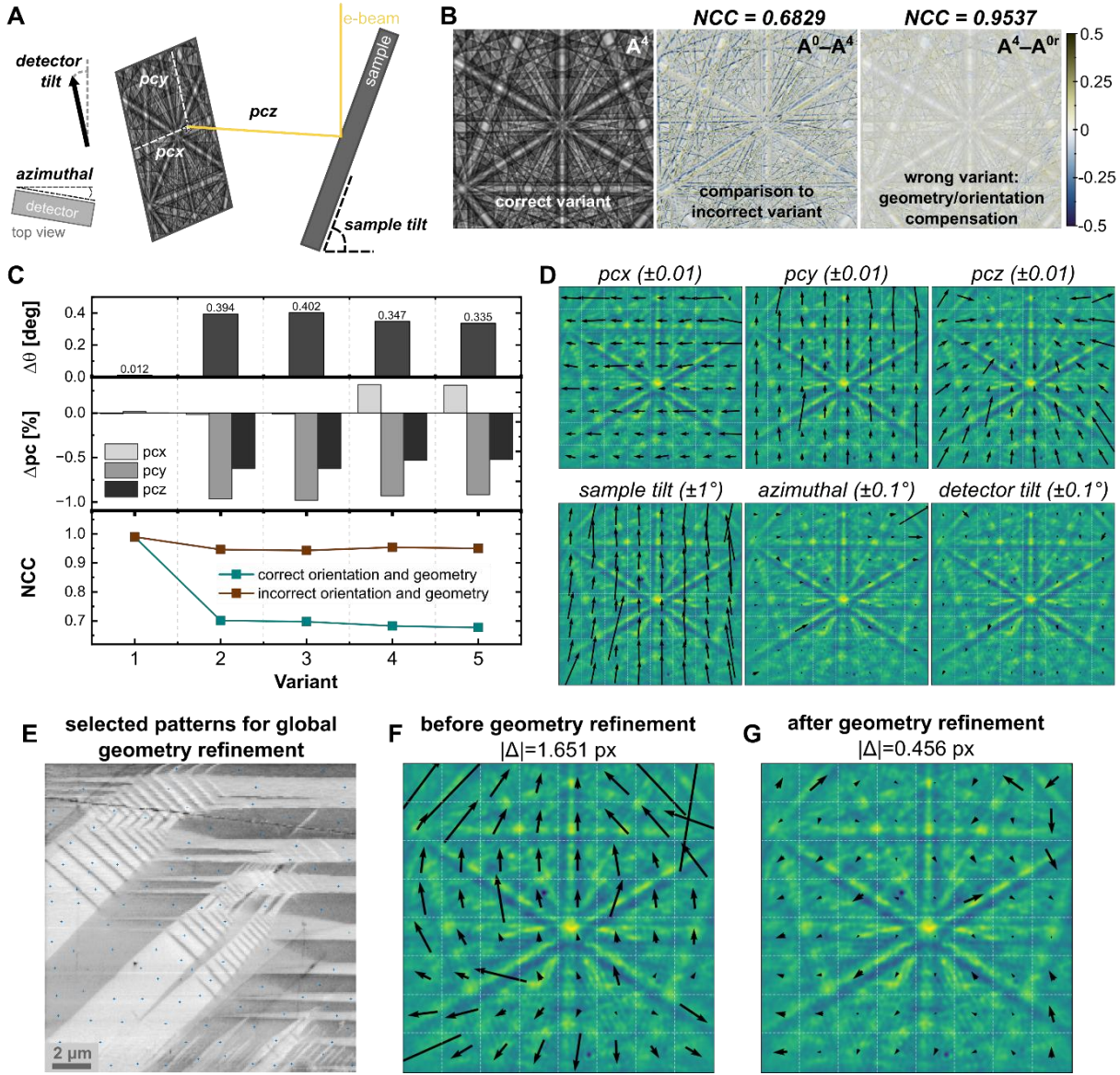

Figure 5: DIC-based global geometry refinement: (A) Schematic of sample-detector geometry showing the six parameters to be optimized; (B) Kikuchi patterns of (from left to right) an example variant (V4 in C), a 90° variant of V4 (V0 – the base orientation) with the difference between patterns overlaid, and a simulated pattern with refined orientation and geometry to match the V4 pattern but with an initial orientation close to V0, with the difference between the pattern of V4 ($A^4$) and the refined solution in V0 space ($A^{0r}$) overlaid; (C) for each variant V1–V5, the change in orientation (misorientation angle $\Delta\theta$) and change in pattern center ($\Delta pc$ in percentage of the respective detector dimension) in the V0 orientation zone to maximize the NCC between each variant pattern and the refined pattern with V0 orientation ($A^{0r}$: an "incorrect" indexing of the variant pattern); (D) displacement fields showing the sensitivity to each geometry parameter; (E) positions on a BTO CI map (denoted by blue crosses) corresponding to the ~100 patterns selected for global geometry refinement; (F) average displacement field (calculated within each ROI across the ~100 patterns) showing the global shifts between simulated and experimental patterns with a total average displacement magnitude of 1.651 px; (G) averaged displacement field after geometry refinement showing significant reduction in local displacements.

There is a fundamental issue with independently optimizing the pattern center for several points within a map containing PS variants: the pointwise changes in orientation and variant cannot be decoupled from the effects of sample-detector geometry. Specifically, we hypothesize that the pattern changes from one variant to another can be approximately duplicated by changes in pattern center and orientation within a different variant's orientation zone. To confirm this hypothesis, we chose a single base orientation of BTO and simulated patterns for this base orientation and its five PS variants. These five PS variant patterns were then taken as a "virtual" experimental dataset with initial orientations set to the base orientation. Simultaneous refinement of the orientation and pattern center using *Kikuchipy's refine_orientation_projection_center* function was carried out with a trust region of 3° in orientation space and 5% detector width/height for the pattern center. Figure 5B shows the refinement results for the 4$^{th}$ variant pattern (a 90° domain variant). The initial NCC score between the base pattern ($\mathbf{A}^0$) and this 4$^{th}$ variant pattern ($\mathbf{A}^4$) is 0.6829 (differences between patterns are highlighted in the middle panel). However, refining the pattern center and orientation within 3° of the base orientation produces a solution in V0 orientation space that has an NCC score of 0.9537 compared to the 4$^{th}$ variant pattern. In this case, the deceivingly high NCC score identifies the wrong variant due to the incorrectly optimized pattern center. Figure 5C shows the NCC results for the other four variants as well as the misorientation angle between the base and refined orientations ($\Delta\theta$) and the changes in pattern center ($\Delta pc$). The NCC scores between the base pattern and variant patterns are shown in teal, representing the correct relationship between variants. Once the orientation and pattern centers have been independently refined in the base orientation space, the NCC scores (brown) between the refined base orientations and variants are significantly higher (>0.94), despite still being the incorrect variant. To achieve this incorrect matching, the pattern centers are changed by <1% of the pattern height/width and the orientation change is <0.41°. This analysis confirms that small deviations in pattern center and orientation can mimic the changes in patterns between variants.

This problem calls for a new solution: one that decouples the pointwise effects of orientation and PS variants from the global geometry calibration. Our method accomplishes this by considering the consistent pattern shifts between experimental and simulated patterns across the map to be signatures of a global sample-geometry problem rather than pointwise effects of orientation. This new method—which we term "*DIC-based global geometry refinement*"—is described in detail in [54], but we briefly describe the key concept here and show results for BTO ferroelectrics.

A selection of patterns is chosen from the map—ideally spanning multiple grains or, in the case of single-crystal ferroelectrics, multiple domains (Figure 5E). For this subset, the initial Hough solution and geometry from the EDAX (or any commercial EBSD) software are used to generate simulated patterns. Each pattern is divided into a grid of ROIs (typically 5-10 divisions per side). Sparse feature tracking is used to compute displacement vectors for each ROI describing the shifts between the experimental and simulated patterns. Then, we average the displacement vectors for each ROI across all patterns. This produces a displacement field representing the consistent shifts between experimental and simulated patterns corresponding to a global mismatch in geometry parameters. Figure 5F shows the average displacement field for the patterns selected from the BTO map (Figure 5E), showing a consistent shift between patterns with an average displacement of 1.651 px between simulated and experimental patterns.

While this average displacement field captures the global geometry mismatch, we aim to connect this discrepancy to specific errors in the geometry parameters. To accomplish this, we compute the displacement sensitivity to changes in each geometry parameter, based on central-difference derivatives of the displacement fields between experimental and simulated patterns. These sensitivities are shown in Figure 5D with the parameter and applied variation shown above each image. Changing *pcx* and *pcy* are predictably associated with horizontal and vertical shifts of the patterns, respectively. Changing *pcz* results in a magnification of the pattern around the *pcx* and *pcy* coordinates. Changing the azimuthal angle is similar to a *pcx* shift, and changing the detector or sample tilts mainly shifts the pattern down or up with some horizontal component towards the edges. We included not only the pattern center parameters in this analysis but also the angular geometry parameters, because the pattern center alone clearly cannot account for the average displacement field of the data. Variation in the sample tilt is possible, as the stage tilt, not the precise sample tilt, is used (which differ due to sample-to-stage mounting tilts). Small variations in the detector angles are also possible due to imprecise installation (~tenths of a degree) or hardware positional shifts after continued usage. Using the displacement field sensitivities of these six geometry parameters, we solve for the linear superposition of parameter changes that maximizes reduction in the average displacement field, and we update the geometry parameters accordingly. Once the orientations are refined, new patterns are simulated, and new average displacement fields are computed. This process of performing DIC-based global geometry refinement and then refining orientations is repeated iteratively, until the NCC score between the experimental and simulated pattern converges (defined as an incremental increase of less than a tolerance of 0.005).

In the case of the BTO map, four iterations were performed to arrive at the final average displacement field in Figure 5G with an average displacement of 0.456 px. The average NCC score improved considerably from 0.4297 to 0.6347. The initial, final, and change in detector parameters are shown in Table 1. Note that due to coupling between parameters, this may not be the only solution. We use this refined geometry for indexing of the BTO maps, whose results (presented in Section 4.1) give evidence that this method enables accurate variant characterization.

**Table 1: Detector parameters before and after DIC-based global geometry refinement**

| geometry parameter | *pcx* | *pcy* | *pcz* | sample tilt [deg] | azimuthal [deg] | detector tilt [deg] | NCC |
|---|---|---|---|---|---|---|---|
| initial value | 0.501 | 0.202 | 0.893 | 69.9 | -2.0 | 10.0 | 0.4297 |
| final value | 0.502 | 0.125 | 0.867 | 73.129 | -1.851 | 9.382 | 0.6347 |

## 3. Confidence index for pseudosymmetric materials

### 3.1. Theoretical landscape of normalized cross correlations

As we have shown in Section 2.2, the normalized cross-correlation between an experimental pattern and patterns of different domain variants can be challengingly similar. To assess the severity of this PS issue as it pertains to ferroelectric domains, we compute the normalized cross-correlation between simulated patterns of different variants. The patterns are simulated using *Kikuchipy* from spherical master patterns which were simulated using *Oxford's MapSweeper* software (details for the crystal parameters, beam energy, and detector parameters are given in the Supplementary Information, Section 5). A 2° equispaced sampling of orientations is defined, and a set of patterns is simulated given these base orientations. Then, the crystal-defined PS operations are applied to the base orientations to result in orientations for the domain variants, which are used to simulate variant pattern sets. The normalized cross-correlation is computed for each pairing of base- and variant-orientation patterns.

We analyze three different ferroelectric materials using this approach: lithium niobate ($LiNbO_3$), lead zirconium titanate (PZT), and barium titanate (BTO). $LiNbO_3$ was chosen, since an EBSD pattern matching technique was previously used to successfully distinguish between the 180° variants of this material [27]. It has a trigonal crystal structure in its ferroelectric phase with a *c/a*-ratio of 2.693. The other two materials, PZT and BTO, are frequently used ferroelectrics that have a tetragonal crystal structure with significantly lower *c/a*-ratios of 1.013 and 1.007, respectively. The NCC theoretical landscapes for these three materials are presented in Figure 6 (A – B). Within the pole figures, the base-orientations are plotted and color-coded by the NCC value of the corresponding 180° variant. The 180° domain variants are most similar, because the only difference is the displacement of the inner cation in the *c*-direction of the unit cell (the unit cell dimensions remain unchanged). With a large *c/a*-ratio, $LiNbO_3$ exhibits low NCC values for 180° domain variant patterns of most orientations. Few orientations have NCC values between 0.05 and 0.25, and only one orientation has a high NCC value of 0.943. The differences between the 180° variant patterns with the maximum and minimum NCC scores are plotted below the pole figures. In contrast to $LiNbO_3$, all orientations of PZT and BTO have high NCC values—above 0.871 and 0.986, respectively. This means that the 180° variant patterns are nearly identical, especially for BTO. The images below the pole figures, which show the differences between the patterns, illustrate the high degree of similarity. With a *c/a*-ratio of 1.007, the unit cell of BTO is nearly cubic, which is reflected in the diffraction patterns. The box plots of Figure 6B show the statistics of the NCC values reported in the pole figures, again demonstrating that the severity of PS for BTO is highest.

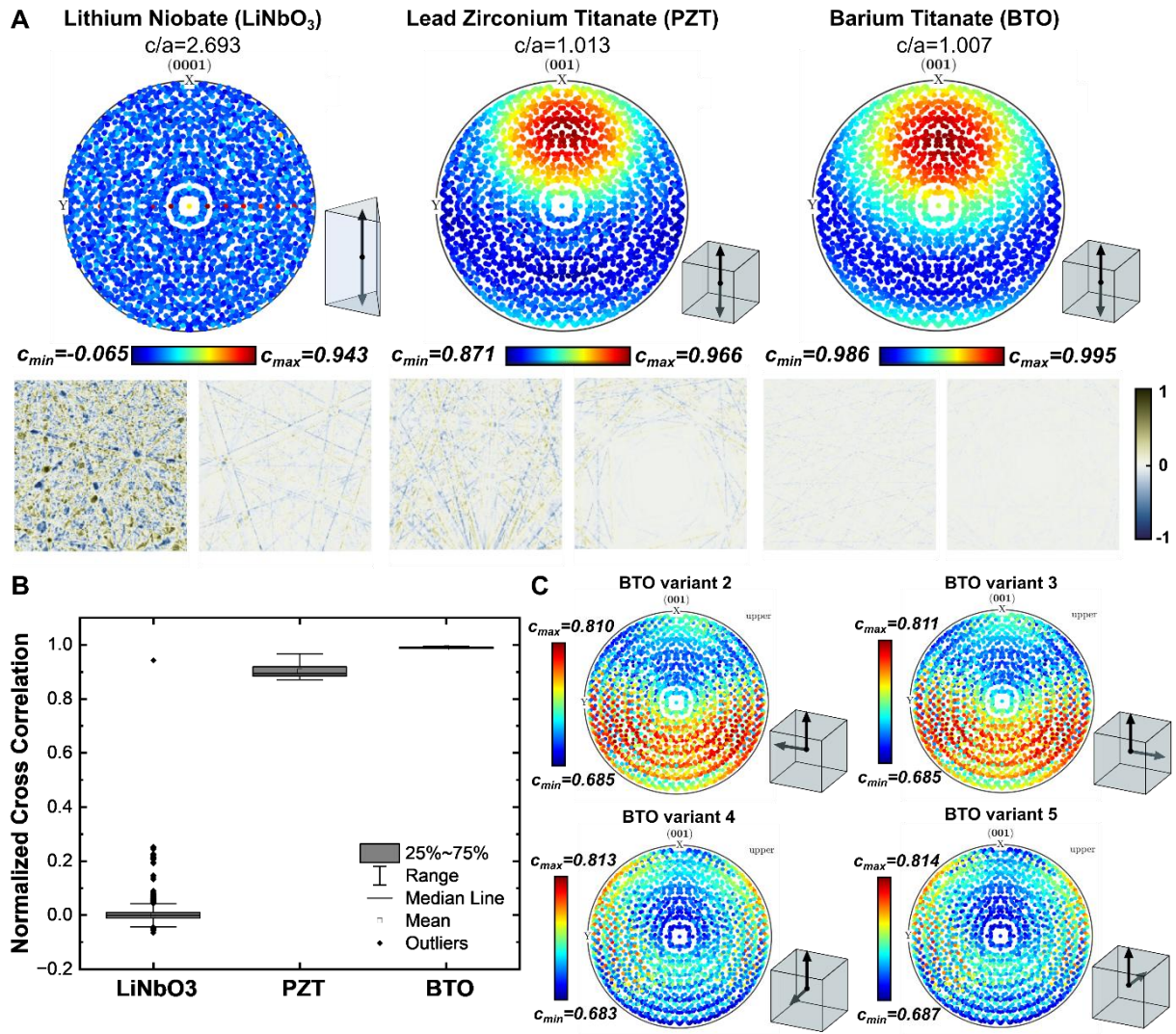

**Figure 6: Theoretical landscape of normalized cross correlations between patterns of different domain variants:** (A) NCC values for 180° domain variants in LiNbO$_3$, PZT, and BTO, shown for different orientations in the pole figures; pattern pairs with the highest ($c_{max}$) and lowest ($c_{min}$) NCC values are shown below the pole figures as differences between the patterns; (B) statistics, represented through box plots, of NCC values reported in the pole figures for the three materials in (A); (C) orientation-dependent NCC values for the four 90° domain variants of BTO.

The pole figures of BTO and PZT in Figure 6A also demonstrate that there is an orientation dependence on pattern similarity between variants. The region in orientation space with high NCC values is in the same area for BTO and PZT, reflecting the similar crystal structure. This region of high NCC is not only dependent on the crystal structure but also on the sample-detector geometry—for example, a different detector tilt would change the diffraction geometry, so that the same orientation would no longer have the same NCC value for a given domain variant. BTO and PZT also have four 90° domain variants. We have performed the same analysis for the 90° variants of BTO (using the same base orientations and detector geometry), whose results are shown in Figure 6C. The PS is less severe for the 90° variants, but the patterns are still highly similar—with NCC values between about 0.68 and 0.81. The pole figures show that, like the 180° variant, there are also certain regions in orientation space with higher NCC values. This analysis would be beneficial to do prior to EBSD of single-crystal materials with PS challenges to ensure the most favorable acquisition geometry is used to distinguish the PS variants of interest.

*3.2. Weighted correlation metric*

With NCC values as high as 0.995 for perfect, simulated patterns of 180° domain variants in BTO, it is nearly impossible to distinguish between polar domains from noisy experimental patterns. However, the simulated patterns provide the key to fully discriminate between these close pseudosymmetric orientations: the local differences in intensities between the two simulated patterns. These differences guide exactly where to focus the pattern matching. Using this logic, we define the *weighted cross-correlation* (WCC) metric

$$c_w(\mathbf{A},\mathbf{B},\mathbf{W}) = \frac{\sum_m \sum_n W_{mn}^2 (A_{mn}-\overline{\mathbf{A}})(B_{mn}-\overline{\mathbf{B}})}{\sqrt{\left[\sum_m \sum_n W_{mn}(A_{mn}-\overline{\mathbf{A}})^2\right]\left[\sum_m \sum_n W_{mn}(B_{mn}-\overline{\mathbf{B}})^2\right]}} \quad (2)$$

where $A_{mn}$ and $B_{mn}$ are the intensities at pixel $(m,n)$, $\overline{\mathbf{A}}$ and $\overline{\mathbf{B}}$ are the image-average intensities, and $W_{mn}$ is the weight at pixel $(m,n)$ computed for images $\mathbf{A}$ and $\mathbf{B}$. The weight matrix $\mathbf{W}^{ij}$ is computed from the simulated pattern $\mathbf{A}^i$ and the simulated variant pattern $\mathbf{A}^j$ as

$$\mathbf{W}^{ij} = \begin{cases} |\mathbf{A}^i - \mathbf{A}^j| & \text{for } i \neq j \\ 1 & \text{for } i = j \end{cases} \quad (3)$$

The benefit of using the WCC metric over the NCC metric is illustrated in Figure 7. The NCC and WCC values are computed from patterns of 180° domain variant pairs in BTO, which were simulated using the same master pattern and detector geometry. The pole figures show uniform distributions of orientations, which are color-coded by the respective NCC or WCC values. Simulated patterns corresponding to the maximum and minimum NCC values are shown along with the computed weights. These weights (calculated for each orientation) are used to compute the WCC, and the orientation-specific WCC values are shown in the pole figure to the right of Figure 7A. The pole figure of WCC values shows the same localization of high values, although this region is smaller compared to the NCC. The primary benefit of the WCC is that the values compared to the NCC are significantly reduced. The ranges go from 0.986 – 0.995 for the NCC to 0.914 – 0.967 for the WCC, reducing the maximum value and extending the range—making it easier to distinguish between 180° domain variants.

The WCC also improves the distinction between 90° domain variants. Figure 7B shows example pole figures of two 90° domain variants color coded by WCC or NCC value. The locations of high correlation values are similar between the two metrics. However, we see a marked reduction in similarity value using the WCC compared to the NCC. Figure 7C shows the average, minimum, and maximum NCC and WCC values for each of the five variants. For all 90° domain variants, the WCC values are less than 1/6 the NCC values and have expanded ranges. This demonstrates the advantage of using the WCC metric over the NCC metric for distinguishing between domain pairs or other close pseudosymmetric orientations.

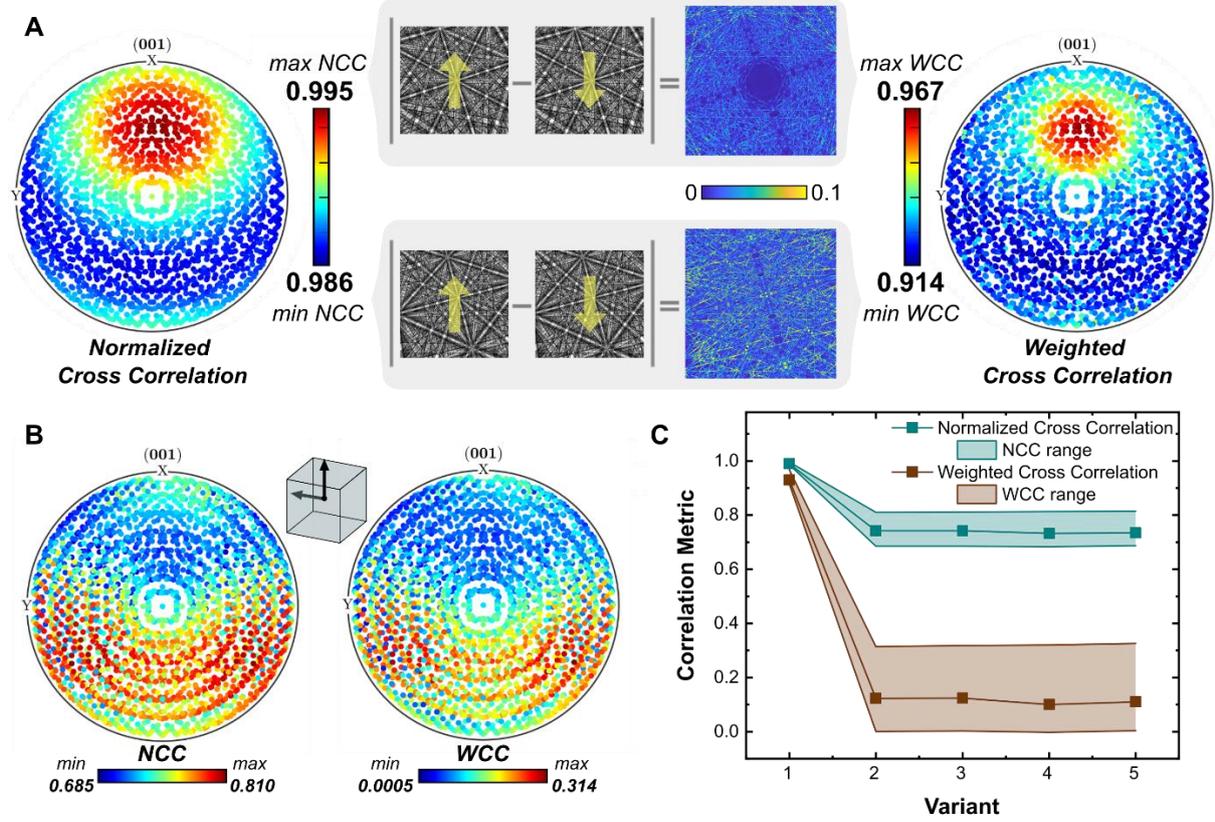

**Figure 7: Comparison of normalized and weighted cross correlations for domain variants in BTO:** (A) Analysis of WCC and NCC for 180° domain variants with example patterns and weight matrices; (B) NCC and WCC values for a 90° domain variant; (C) averages and ranges of NCC and WCC values for each variant (variant 2 is the 180° domain variant and 3-5 are the 90° domain variants).

### 3.3. Pseudo-Symmetry confidence index ($CI_{PS}$)

The previous analysis used the WCC to compare two simulated patterns of different variants. This method must be adapted to use with experimental patterns, where the correct variant solution is uncertain and must be confidently determined. Here, we introduce a new confidence index, which leverages the WCC metric to maximize distinguishability between Kikuchi patterns of pseudosymmetric variants. The new confidence index assesses the match between an experimental pattern and simulated patterns of **all** PS variants. First, the indexed orientation for each pixel is used to simulate a pattern corresponding to this *base* orientation. These *base* orientations are from an indexed EBSD map—either from Hough indexing, spherical indexing, or refinement/dictionary indexing. The simulated pattern of the *base* orientation is denoted by $\mathbf{A}^0$. Then, rotation operations describing the PS variants are applied to the *base* orientation and patterns are simulated with these *variant* orientations (these patterns are denoted by $\mathbf{A}^j$). For example, the tetragonal ferroelectrics studied in this work have 6 possible polarization directions in the unit cell corresponding to different PS variants: the *base* orientation, a 180° rotation about the crystal x-axis, and four others described by ±90° rotations about the crystal *x*- and *y*-axes. Thus, for each *base* orientation, six patterns are simulated, and weight matrices are computed from pairs of simulated patterns as given by Equation (3). For the tetragonal ferroelectrics with five variants (not including the *base* orientation), there are five weight matrices $\mathbf{W}^{0j}$, where $j = \{1,\ldots,5\}$ denotes the variant number. These weights are used to compute the theoretical weighted cross correlation for each variant, using Equation (2):

$$\xi_j^t = c_w\left(\mathbf{A}^0, \mathbf{A}^j, \mathbf{W}^{0j}\right). \tag{4}$$

Similarly, a normalized weighted cross-correlation is computed for the experimental pattern **B** compared to each variant pattern as

$$\xi_j^e = \frac{c_w\left(\mathbf{B}, \mathbf{A}^j, \mathbf{W}^{0j}\right)}{c_w\left(\mathbf{B}, \mathbf{A}^0, \mathbf{W}^{0j}\right)}, \tag{5}$$

where we normalize by the base orientation values, so that $\xi_0^t = \xi_0^e = 1$. Figure 8B shows six plots of $\xi^t$ and $\xi^e$ values per variant. Ideally, the $\xi^t$ and $\xi^e$ curves should be identical, *indicating that the experimental pattern's similarity to each variant pattern is theoretically consistent*. Numerically, this similarity is measured by the mean absolute difference, which in this context we call the *PS variant check*:

$$\eta = \frac{1}{N}\sum_v \left|\xi_v^t - \xi_v^e\right|. \tag{6}$$

For perfectly matched patterns, considering all PS variants, the value of $\eta$ is zero; a large $\eta$ demonstrates a poor match. Consequently, we use the $\eta$ value as a penalty, defining the *PS confidence index* as:

$$CI_{PS} = c - \eta, \tag{7}$$

where *c* is the normalized cross correlation between the experimental pattern and simulated pattern of the *base* orientation.

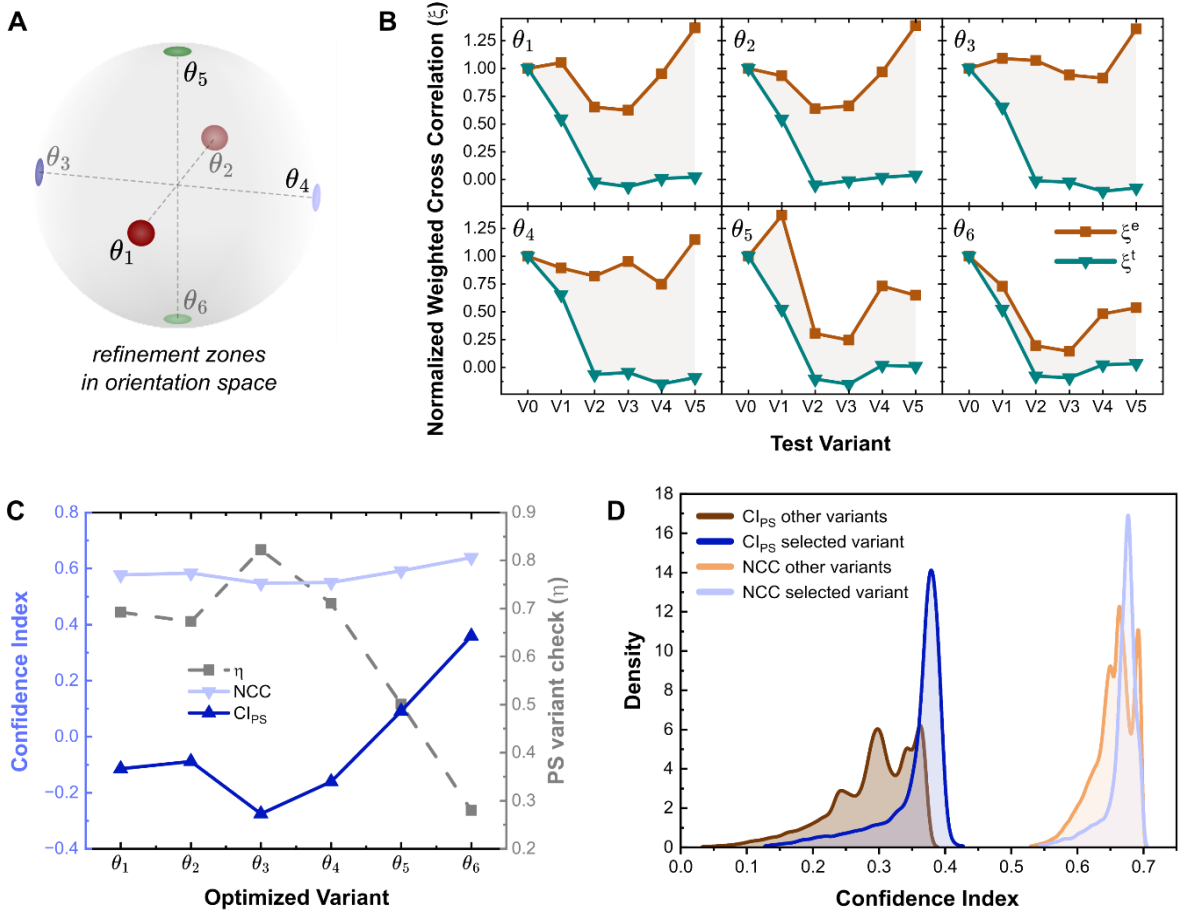

**Figure 8: Pseudosymmetry Confidence Index and Reindexing:** (A) Illustration of the orientation space with six refinement zones corresponding to the six domain variants; (B) theoretical and experimental normalized weighted cross-correlation curves for each optimized variant orientation $(\theta_i)$; (C) confidence index metrics for each optimized variant orientation (same point as in (B)); (D) density plots of NCC and $CI_{PS}$ values for the selected variants and the other five unselected variants for an EBSD map taken on BTO.

*3.4. Implementation of new reindexing scheme*

In principle, this PS confidence index could be used as the objective in an orientation optimization algorithm. Since the patterns must be simulated six times and many more computations performed for each optimization step, the required computational time and resources become infeasible (reindexing speed of 0.06 patterns/second). However, we can use the PS confidence index to select the correct variant out of six independently optimized orientations.

The refinement procedure is carried out by applying *Kikuchipy's refine_orientation* function *[41]*, using the *SciPy* Powell minimization algorithm with a trust region of 2 degrees. The patterns were processed using the Bayesian optimization routine described in Section 2.3, and then the PSS-NPA algorithm described in Section 2.4 was applied to further improve the pattern quality and reduce noise. Geometry calibration was performed using the DIC-based geometry optimization approach outlined in Section 2.5. For the single-crystal BTO EBSD data, ~100 points were manually selected from a map to ensure representation from each domain type for the global geometry calibration. For the polycrystalline

PZT data, a 10x10 equally spaced grid of points was selected for the geometry calibration, which should capture multiple domain variants and crystal orientations—reducing the possibility of orientation contributions to the global geometry calibration. After geometry calibration on the subset of patterns, *Kikuchipy's extrapolate_pc* function was used to define a detector with spatially varying pattern centers for each point in the full map in accordance with the scan geometry and sample/detector tilts. The master patterns for BTO and PZT were simulated using *Oxford's MapSweeper* with the crystal and e-beam parameters defined in the Supplementary Information, Section 5. Finally, the original Hough-indexed orientations are used as a starting point. We then apply the PS relations for tetragonal ferroelectrics to arrive at five additional sets of starting orientations. Each orientation set is refined within a 2-degree trust window. These six refinement zones are represented in the schematic of Figure 8A.

This 6-step refinement yields six independently refined orientation sets covering all possible domain variants. We apply our PS confidence index to choose the correct domain variant for each point. For each optimized variant orientation $\theta_i$, the PS confidence index metrics defined in Section 3.3 are computed. The optimized variant orientation is taken as the *base* orientation, and the PS operations applied to yield the five *variant* orientations. Simulated patterns of the *base* orientation $\mathbf{A}^0$ and five variant orientations $\mathbf{A}^j$ are compared to one another and the experimental pattern to produce the normalized weighted cross-correlation curves ($\xi$; example curves in Figure 8B). From these curves, the PS variant check metric $\eta$ is computed. Finally, the PS confidence index is computed. The optimized variant $\theta_i$ with the highest $CI_{PS}$ is taken to be the correct orientation.

To illustrate this process and the benefit of using the $CI_{PS}$ over the NCC metric, we select one point and plot the correlation metric values in Figure 8B–C. Figure 8B shows the experimental and theoretical normalized weighted cross correlation values for each "test" variant $j$, given each optimized variant orientation $\theta_i$. Comparing the $\xi^t$ and $\xi^e$ curves for each optimized variant, it is clear that $\theta_6$ has the lowest value of $\eta$ (or the smallest mean absolute difference between the two curves). This is apparent in Figure 8C, where the $\eta$, $CI_{PS}$, and NCC values are plotted for each optimized variant orientation. $\theta_6$ has the lowest $\eta$ and the highest NCC and $CI_{PS}$ values and is therefore selected as the correct variant. For this point, the variant with the highest NCC is in fact the correct variant, but this is not always the case. (Since the NCC is not as sensitive as our $CI_{PS}$ to small changes between variant patterns, the correct variant could have the highest $CI_{PS}$ but not the highest NCC score.) The NCC values for all variants are significantly closer than the broad range covered by the $CI_{PS}$ values. This enhanced discernability between variants is illustrated in Figure 8D. The NCC and $CI_{PS}$ values for the six independently refined orientation sets of a single-crystal BTO EBSD map are separated by the values of the selected variants and all the other variants. The density plots for each of these four confidence metric groups reveal narrow windows of NCC values with a high degree of overlap between the selected variant and others. In contrast, the distribution of $CI_{PS}$ values of selected variants is narrow and distinct from the wide distribution of $CI_{PS}$ values of the other unselected variants. This analysis demonstrates that the PS confidence index proposed in this work is superior at distinguishing between close PS variants compared to the NCC metric.

The benefits of the new reindexing scheme come with high computational costs compared to existing schemes. For example, with parallel processing it took ~24 hrs using 20 MPI tasks to reindex the PZT map presented in this work, compared to the ~4 hrs it took to acquire the map and orientation data using Hough indexing. Larger memory resources are also required to store the Kikuchi pattern datasets (~38 GB in the case of the PZT map). Therefore, this method is best suited for challenging applications such as materials with high pseudosymmetry.

## 4. Case studies on ferroelectric materials

### 4.1. Single-crystal BTO

The final reindexing results for the single-crystal BTO maps are shown in Figure 9. Prior to reindexing, the pattern processing, PSS-NPA, and geometry calibration routines were applied, as described in Sections 2.3–2.5; the specific parameters associated with these methods are listed in the Supplementary Information, Section 5. Using *Kikuchipy*, six orientation refinement passes were completed—one for each orientation zone determined by the five tetragonal ferroelectric PS operations applied to the original Hough orientations. From these six orientation sets, patterns were simulated and the $CI_{PS}$ metrics computed. The orientation with the maximum $CI_{PS}$ was selected for each pixel and added to the final refined map. The domain orientations of the final refined map are shown in Figure 9B, colored by the unit cell color map shown between Figure 9A and B. The domains are successfully identified—illustrated by clear divisions between different colored domains with limited noise. The domain laminate pattern towards the right of the scan has a light green–pink alternating order, which is consistent with the PFM results and shows a compatible 90° laminate. Towards the top of the scan, there is a transition within a pink domain to a red domain: this transition marks a compatible 180° domain wall, which is also consistent with the PFM scan. On the left side of the scan, there is another large laminate pattern, which consists of an out-of-plane domain orientation (light blue—into the page) and an in-plane domain orientation (light green, top-left direction). These orientations are consistent with the PFM scan and suggest the wall is at a 45° angle in the out-of-plane dimension to ensure compatibility. These results prove that the EBSD reindexing method introduced in this paper can accurately distinguish between the domain variants of BTO—a case with extreme PS challenges.

We can also use the confidence index metrics to filter out those regions that may be incorrectly indexed. Figure 9C shows the $CI_{PS}$ map for the final selected orientations. The lowest $CI_{PS}$ values are at the domain boundaries and in the fine laminates where pattern convolution between different domains may occur. Elsewhere, the $CI_{PS}$ value is fairly constant across the map at a value of around 0.4, which shows the correct variant is consistently chosen. The $CI_{PS}$ values could be increased, if the pattern quality was better (improving the NCC score) or if the $CI_{PS}$ metric was used in the optimization algorithm. However, as we previously mentioned, computing the $CI_{PS}$ for every iteration of the optimization is computationally expensive. Another useful metric is the *cross-variant margin* (CVM), which is the difference between the $CI_{PS}$ values of the two top domain variants. The CVM provides a measure of discrimination between domain variants: the higher the value, the more confidence we have that the selected variant is correct. The CVM map is plotted in Figure 9E, using a range of 0 to 0.1. These values can be interpreted in terms of the theoretical WCC or NCC values for 180° variants, as shown in Figure 7, since the variant with the second highest $CI_{PS}$ is likely the 180° variant. Considering the NCC values for 180° variants, a theoretical range of the CVM (when using the NCC as the confidence index) is from 0.005 to 0.014, representing the maximum possible difference between perfect patterns. In reality, experimental patterns never achieve such CVM values in the NCC scores. Our $CI_{PS}$ CVM values are one to two orders of magnitude higher than the best CVM values when using the NCC metric—again demonstrating our method's superior distinguishability between PS variants. In Figure 7D, only the polarization orientations of pixels with $CI_{PS}$ > 0.25 and CVM > 0.01 are plotted, filtering out any pixels that may have been incorrectly indexed. The incorrectly indexed pixels' (pink pixels within the light blue domains and the blue pixels at the boundaries of the pink domains) disappearance confirms that this is an effective filtering strategy and that both metrics robustly define the indexing confidence.

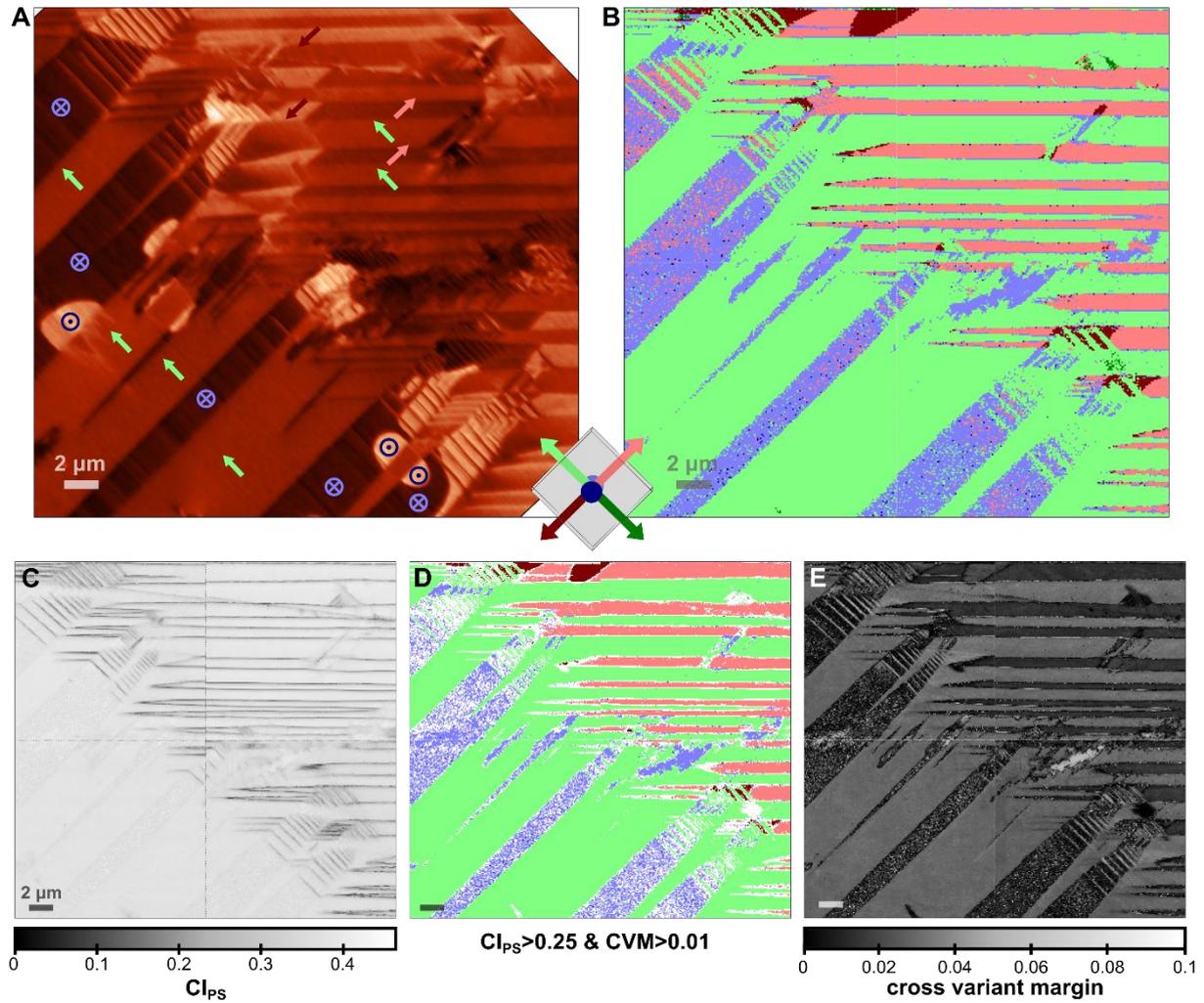

**Figure 9: CI$_{PS}$ reindexing of single crystal BTO EBSD maps:** (A) Out-of-plane PFM signal identifying correct polarization orientations (identified through vector PFM analysis in SI, Section 2); (B) domain orientations from CI$_{PS}$ based reindexing of EBSD maps; (C) CI$_{PS}$ map; (D) same map as B with pixels having a CI$_{PS}$<0.25 and a CVM<0.01 removed; (E) cross variant margin map (difference in CI$_{PS}$ between the top two domain variant candidates).

*4.2. Polycrystalline PZT*

Having validated our new $CI_{PS}$ reindexing method with known domain orientations within a BTO single crystal, we test the method on a polycrystalline sample. Currently unachievable with other methods and of significant potential for the study of domain microstructures in polycrystalline ferroelectrics, we demonstrate the accurate, spatially resolved identification of the polarization directions and crystal orientations in polycrystalline PZT. PZT has a slightly larger *c*/*a* ratio than BTO, so discriminating between the domain variants is slightly easier but still difficult when using the NCC metric, where Kikuchi patterns of 180° variants have NCC scores up to 0.966 (see Figure 6). Due to the NCC dependence on crystal orientation, it may be easier to distinguish between domain variants in some grains and harder in others. Figure 10A shows the grain orientations with a cubic IPF color code (grains identified by increasing the crystal symmetry from tetragonal to cubic and applying the *calcGrains* function in *MTEX* [55]) and the grain boundaries color-coded by misorientation angle. To the original

Hough results and patterns, we have applied the same the pattern processing, PSS-NPA, and geometry calibration routines described in Sections 2.3–2.5 with the corresponding parameters listed in the Supplementary Information, Section 5.

After the six individual refinements and selection of the best variant based on the $CI_{PS}$, the final selected domain orientations are plotted in Figure 10B. Note that each grain has a different polarization direction color coding, as defined by the unit cell schematics within each grain of Figure 10A. The axes of the unit cell are colored by which grain-frame directions are closest to the specimen frame directions. For example, the axis pointing "most out-of-the-page" is always colored dark green, the axis pointing "most up" is always colored dark blue, and the axis pointing "most left" is always colored red. Using this color scheme, similarities in polarization directions between grains are visually apparent, although the same color in different grains does not correspond to the exact same direction. Quivers are overlaid on the color plot to make this more obvious.

Judging by the clarity of domain boundaries and absence of noise (Figure 10B), the indexing is successful. Looking closely at the polarization directions and the domain boundaries (intersections of walls with the surface), it also appears that all domain walls are compatible. For example, the grain at the top right contains a pink–light-blue laminate pattern, which has 90° domain pairs with the boundaries at an orientation consistent with being inclined 45°, as is typical for compatible 90° domains. Note that the domain walls are not simple extrusions of the domain boundaries in the out-of-plane dimension but can be defined by crystallographic and compatibility conditions. Another obvious example of compatible domains is within the laminate of the largest central grain: the boundary between the red and pink domains is aligned with the polarization directions, making a compatible wall between 180° domains.

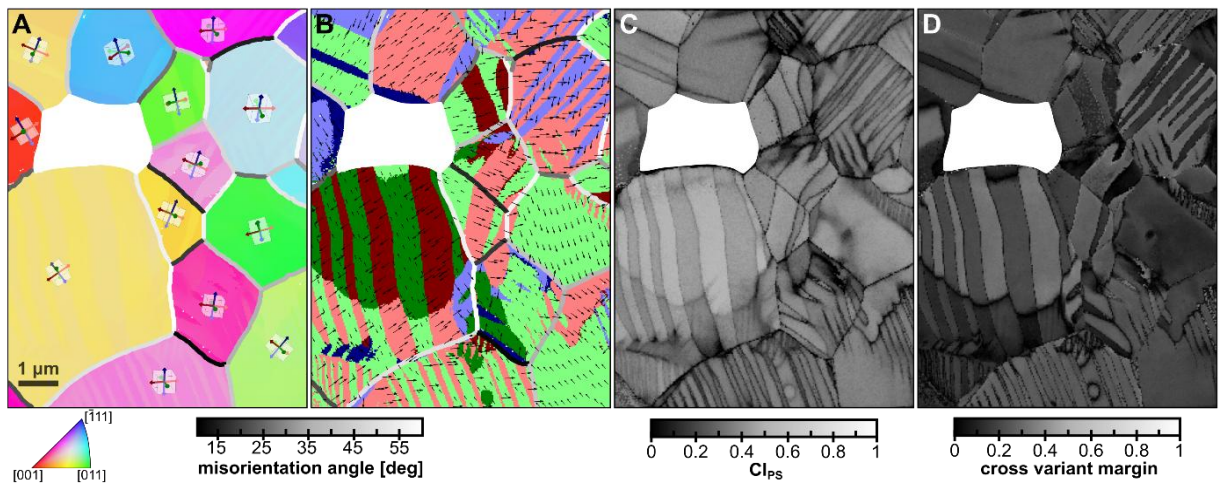

**Figure 10: $CI_{PS}$ reindexing of a polycrystal PZT EBSD map:** (A) Cubic IPF coloring of orientation map showing grain orientations and boundaries with unit cells showing grain-frame polarization direction color keys for B; (B) domains colored by the grain-frame polarization orientations with grain boundaries in A and B colored by the misorientation angle color map shown below the figures; (C) $CI_{PS}$ map; (D) cross variant margin map (difference in $CI_{PS}$ between the top two domain variant candidates).

The indexing success can be quantified using the $CI_{PS}$ and CVM metrics (Figure 10 (C–D)). Both maps clearly show the domain microstructure, with low values (dark) at domain boundaries. The $CI_{PS}$ values of the domain interiors are typically high, usually between 0.35 and 0.55. The CVM typically ranges from ~0.1 to 0.3 within the domains, which is significantly larger than the theoretical CVM when using the NCC, which ranges 0.03 to 0.13. These results indicate that the indexing results are accurate.

The domain orientation plot in Figure 10B highlights the potential of this new technique in the study of ferroelectric polycrystals. A recent study examined domain continuity across a small sampling of grain boundaries, using orientations determined from selected area electron diffraction patterns [56]. With our new EBSD reindexing method, we can study the conditions for domain continuity across grain boundaries at a more quantitative and statistical level, capturing the full spatially resolved domain microstructures within different grains. This small map in Figure 10B reveals domain continuity across almost every grain boundary, sometimes even traversing multiple grain boundaries (e.g., the pink domains across the central grains). We also see fine laminate structures forming along grain boundaries and 180° curved domains intersecting 90° laminates. These preliminary observations are only the starting point and hint at what quantitative information can be extracted from experimentally imaged domain structures in polycrystalline ferroelectrics, using this method.

## 5. Conclusions and outlook

We have presented and validated an EBSD reindexing method, which accurately identifies spatially resolved domain polarization directions in ferroelectric materials. Applied to tetragonal ferroelectrics (specifically, single-crystal BTO and polycrystalline PZT), which have six possible polarization variants, we show that the exact polarization direction can be uniquely determined. As BTO has a nearly cubic crystal structure, Kikuchi patterns of 180° domain variants are nearly identical and cannot be accurately indexed using state-of-the-art EBSD pattern-matching approaches. Our new PS confidence index greatly improves distinguishability between patterns, enabling the correct domain variant to be accurately identified (which is especially challenging for 180° domain variants). We have proven that our method accurately determines the polarization direction by comparing the EBSD results to PFM scans of the same area. The demonstrated success with BTO implies that domain identification in virtually any ferroelectric material should be possible with this technique.

This new reindexing method is not only suitable for ferroelectrics but generally to any material which exhibits pseudosymmetry (PS). We have presented a method to assess the severity of PS issues for a given material by simulating patterns of all variants for different orientations and visualizing the theoretical NCC landscapes on pole figures. We have shown that the WCC is a more discriminatory metric, providing wider margins between different variants. Finally, we have incorporated this WCC metric into a new PS confidence index and utilized this metric for selecting the correct PS variant. Our study may initiate a paradigm shift in EBSD pattern matching approaches, by realizing that the NCC is not always the best metric to use. We do not claim that the $CI_{PS}$ metric used in this study is the best for every material and application (there may even be a better metric for this study), but we encourage the exploration of different confidence index formulations to improve indexing accuracy. However, given the increased computational time and memory requirements, we advise performing preliminary analysis on small scans to test the traditional EBSD analysis methods before employing the computationally intensive methods presented in this paper. Additionally, this method is typically not necessary for most materials analyzed by EBSD and should only be applied in situations where traditional EBSD analysis methods fail.

In addition to the new PS confidence index and method to assess PS issues, we have created new pre-processing methods, which improve indexing success. The first is a method for automatic, quantitative determination of the best pattern processing parameters using Bayesian optimization with Gaussian processes. This method finds the best pattern processing parameters to optimize the pattern quality and match (NCC score)—removing the guess work out of pattern processing. The second method is a new PS-sensitive neighbor pattern averaging (PSS-NPA) method, which is similar to the NLPAR algorithm but is more selective about which patterns to include in the averaging. We show that this method is superior to the NLPAR algorithm for materials with close PS, since nearly identical patterns of different PS variants are typically included in the NLPAR averaging scheme. We have also demonstrated successful usage of our DIC-based global geometry refinement algorithm, which decouples the orientation-dependence from the optimization of global sample-detector geometry parameters. Again, these methods are not only applicable to EBSD on ferroelectrics but widely applicable to EBSD of challenging PS materials. With the successful application to polycrystalline PZT, our new reindexing method opens the door to quantitatively studying domain microstructures in polycrystalline ferroelectrics. There are many unresolved questions related to domain formation and evolution, including the effects of crystal orientation, grain boundary misorientation, grain morphology and sizes, pores and defects, and the effect of applied electric field and stress on domain microstructures. The presented method provides full spatially resolved microstructural information, including the grain and domain orientations, which is currently unachievable with other methods such as PFM, PLM, and XRD.

**Data and Code Availability**

Upon publication, all EBSD data will be made publicly available, hosted on ETH Zurich's Research Collection. All developed code—including the pattern processing optimization, PSS-NPA, DIC-based global geometry refinement, and reindexing with the PS confidence index algorithms—will be made publicly available on GitHub.

**Acknowledgements**

The authors gratefully acknowledge the support and assistance from Mathieu Brodmann, Hsu-Cheng Cheng, Christian Franck, Vignesh Kannan, and Karsten Kunze, as well as the ScopeM center at ETH Zurich. The authors gratefully acknowledge the financial support from the Swiss National Science Foundation (SNSF) under projects 212643 and 236413.

# Supplementary Information

# for

# Ferroelectric polarization mapping through pseudosymmetry-sensitive EBSD reindexing

Claire Griesbach[1], Tizian Scharsach[2], Morgan Trassin[2], Dennis M. Kochmann[1]

[1] Mechanics & Materials Laboratory, Department of Mechanical and Process Engineering, ETH Zürich, 8092, Zürich, Switzerland

[2] NEAT Lab, Department of Materials, Zürich, 8093, ETH Zürich, Switzerland

1. Sample preparation and SEM/EBSD parameters

The samples in this work were procured commercially: the (001) BTO single crystals were purchased from Princeton Scientific and the PZT polycrystals were purchased from PI Ceramic GmbH. Samples were ground and polished using an Allied High Tech MultiPrep mechanical polisher, typically following the steps in Table 1.

| Step | Grinding/polishing medium | Pad | Lubricant | Load (0-5) | RPM | Approx. time [min] |
|---|---|---|---|---|---|---|
| 1 | 600 grit Allied SiC paper | | water | 4 | 75 | 3 |
| 2 | 1200 grit Allied SiC paper | | Water | 2 | 75 | 5 |
| 3 | 2500 grit Allied SiC paper | | Water | 2 | 75 | 3 |
| 4 | 1 μm Allied diamond suspension | Allied red magnetic pad | Struers alcohol-based blue lubricant | 2 | 50 | 25 |
| 5 | 0.1 μm Allied diamond suspension | Allied red magnetic pad | Struers alcohol-based blue lubricant | 2 | 50 | 18 |
| 6 | 0.04 μm Struers colloidal silica suspension | Allied black chem-pol pad | | 2 | 50 | 15 |

Table 1: Grinding and polishing steps for sample preparation

Samples were then sonicated in acetone to remove them from the polishing chuck, then sonicated three additional times in acetone, isopropanol, and deionized water before mounting on an SEM stub with carbon paint. The samples were plasma-cleaned for 10 min prior to SEM imaging. This rigorous polishing and cleaning procedure ensures optimal contrast between domains, rather than contrast from topographical features dominating.

SEM imaging and EBSD analysis were performed in a TFS Quanta 200F SEM equipped with a EDAX Hikari EBSD detector. The parameters in Table 2 were typically used for SEM imaging to avoid charging-induced domain evolution:

| Material | Vacuum pressure [Pa] | Acc. Voltage [kV] | Beam spot size | Aperture | WD [mm] | Resolution | Dwell time [μs] | Line integration |
|---|---|---|---|---|---|---|---|---|
| BTO single crystal | 30 | 20 | 5 | 5 | 10 | 2048x1768 | 30 | 2 |
| PZT polycrystal | 60 | 5 | 4 | 5 | 10 | 2048x1768 | 30 | 2 |

Table 2: Typical parameters used for SEM imaging of the BTO and PZT samples

These parameters were optimized to maximize domain contrast without causing charging-induced domain evolution. After finding a region of interest and recording the coordinates, SEM images were taken at multiple tilt angles to maximize contrast for different domain orientations and reconstruct the full domain microstructure in a hyperspectral image (as shown in Figure 1 of the manuscript). After SEM imaging and prior to EBSD analysis, the sample was coated with 2 nm carbon. The nominal EBSD parameters used for each sample are provided in Table 3.

| Material | Vacuum pressure [Pa] | Acc. Voltage [kV] | Beam spot size | Aperture | Gain | Exposure | Step size [nm] |
|---|---|---|---|---|---|---|---|
| BTO single crystal | 30 | 25 | 5 | 5 | 400 | 450 | 80 |
| PZT polycrystal | 30 | 15 | 5 | 5 | 400 | 400 | 30 |

Table 3: Parameters used for EBSD analysis of the BTO and PZT samples

All samples were tilted by 70° for EBSD analysis and no pattern binning was applied (1x1) to retain intensity details. Samples were brought to a working distance of 15–18 mm, until illumination on the detector was centered and symmetrical under the appropriate gain and exposure.

2. Piezoresponse Force Microscopy (PFM)

Piezoresponse force microscopy (PFM) measurements were conducted using an NT-MDT NTEGRA scanning probe microscope in combination with two external Stanford Research SR830 lock-in amplifiers and an external bias-voltage amplifier. The use of two lock-in amplifiers allows for a simultaneous acquisition of vertical and lateral PFM (VPFM and LPFM, respectively) signals. The converse piezoresponse of the sample was induced by applying a 12.5 V AC modulation to a conductive Pt-coated HQ:NSC35/Pt AFM-tip from MikroMasch at a frequency of 76 kHz. VPFM and LPFM were recorded in Cartesian coordinates to ensure the comprehensive capture of polarization vector information [1]. The PFM was first calibrated on periodically poled LiNbO3 and subsequently fine-tuned on the BTO sample, compensating for phase offset ($\phi$) in each lock-in amplifier and cross-talk between vertical and lateral signals induced by the hardware [2].

The polarization orientations were determined from the four PFM signals in Figure S1 (A-D), using the following reasoning. Since the cantilever was roughly aligned along the crystal orientation (known with respect to the domain microstructure from EBSD), we can reason out the in-plane polarization components from the LPFM signals. The sample was rotated by 90°, so that both in-plane orientations could be probed. Since the piezoresponse is directly related to the polarization orientation, the twisting of the cantilever—which is visible in the intensity of the lateral signals—corresponds to the in-plane horizontal component of the polarization. Similarly, the VPFM signal corresponds to an out-of-plane piezoresponse from the out-of-plane polarization component. Thus, since the cantilever is aligned with the crystal orientation and we only have six possible polarization orientations in single-crystal BTO, regions with the highest and lowest intensities in each signal can be attributed to the respective polarization orientations.

Looking first at the Orientation 1 LPFM signal (Figure S1 (A)), we observe that the laminate pattern on the left side of the scan has domains with the highest and lowest intensity in the scan: therefore, we can logically assign these dark and light regions to have left and right polarization directions, respectively. We can similarly assign the dark and light regions in the Orientation 2 LPFM scan (Figure S1 (C)). The out-of-plane signals should in general show similar information between the two scan orientations, although there are some in-plane contributions visible in the out-of-plane channel due to the similar cantilever deformation modes from in-plane polarization components parallel to the cantilever. Nevertheless, we can assign out-of-plane polarization directions to the domain regions that are consistently dark or bright in the VPFM signals (Figure S1 (B, D)). Using this logic, the final polarization directions are indicated in Figure S1 (E).

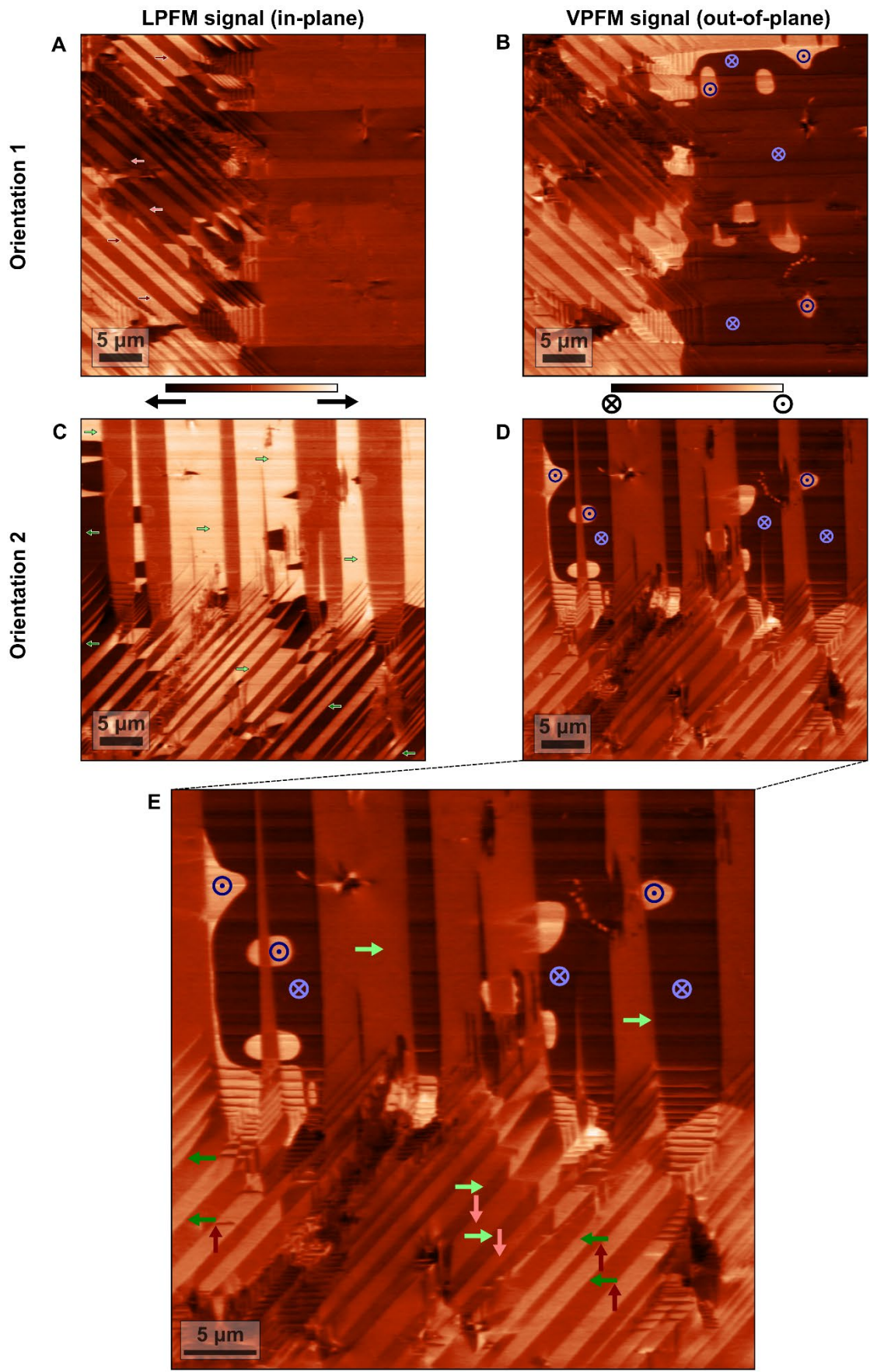

**Figure S1: PFM signals labeled with polarization directions:** (A) Orientation 1 LPFM signal, enabling identification of two in-plane polarization directions; (B) Orientation 1 VPFM signal, enabling identification of the two out-of-plane polarization

directions in conjunction with the Orientation 2 VPFM signal in **(D)**; (C) Orientation 2 LPFM signal, enabling identification of the other two in-plane polarization directions; (D) Orientation 2 VPFM signal, also used to identify the out-of-plane domains; (E) Orientation 2 VPFM signal with all polarization directions identified from (A-D) overlaid.

## 3. Optimization of pattern processing steps

| Dim. | Processing step | Parameter | Type | Values/Range |
|---|---|---|---|---|
| 1 | Dynamic background subtraction | standard deviation | integer | 4 – 40 |
| 2 |  | truncation | integer | 2 – 10 |
| 3 | Adaptive histogram equalization | switch | categorical | True/False |
| 4 |  | kernel | categorical | 48, 64, 80, 96, 112, 128, 256 |
| 5 |  | clip | categorical | logspace(1e-4, 5e-3) |
| 6 |  | nbins | categorical | 128, 256, 512 |
| 7 | FFT filter | high-pass cutoff | integer | 1 – 7 |
| 8 |  | low-pass cutoff | integer | 50 – 100 |

**Table 4: Pattern processing parameters used in Bayesian optimization**

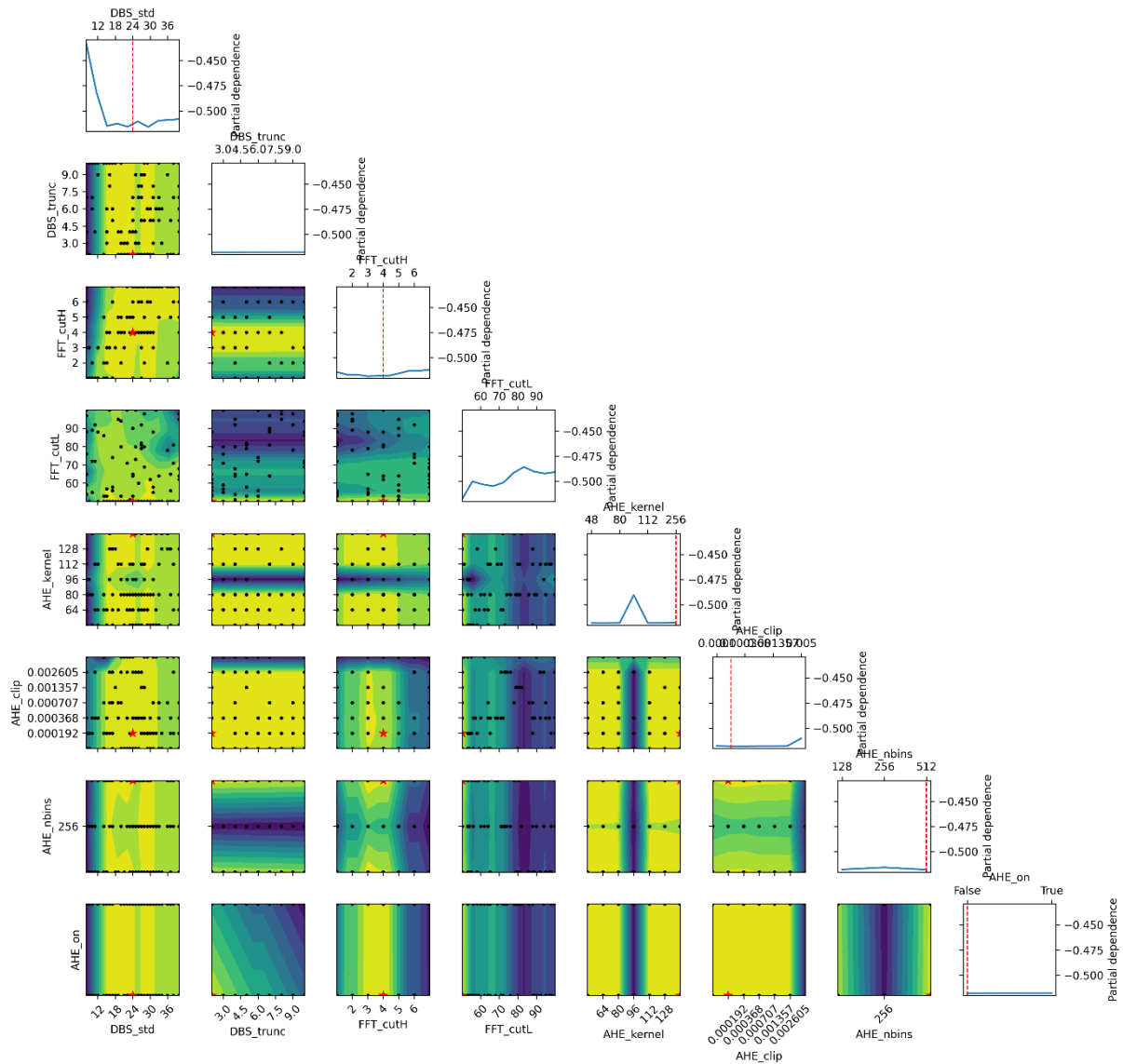

**Figure S2: Bayesian optimization of pattern processing parameters**: each color plot shows the optimization landscape between two parameters.

## 4. Pseudo-Symmetry-Sensitive Neighbor Pattern Averaging (PSS-NPA)

In the manuscript, we introduced the new Pseudo-Symmetry-Sensitive Neighbor Pattern Averaging (PSS-NPA) scheme and illustrated its superiority over the NLPAR technique in retaining division between neighboring patterns with close pseudosymmetries. The detailed algorithm behind the method is as follows.

First, we define the patterns in the map as a 4D array: $\mathbf{B}_{i,j}(y,x)$, where $i$ and $j$ are row and column indices corresponding to the pattern's location in the EBSD map, and $y$ and $x$ are the pattern pixel coordinates. Given a square grid, we use a Manhattan search radius $r$ to define the neighborhood as $\mathcal{N}_r = \{(\Delta i, \Delta j) : 0 < |\Delta i| + |\Delta j| \leq r\}$ and the neighbors of the central pixel $(i,j)$ as $N_{i,j} = \{(i+\Delta i, j+\Delta j) : (\Delta i, \Delta j) \in \mathcal{N}_r\}$.

For each neighbor $k$, we compute the normalized cross-correlation (NCC) between the central pixel and any of its neighbors, as defined by Equation (1) in the manuscript: $c_k\left(\mathbf{B}_{i,j}, \mathbf{B}_{i+\Delta i_{(k)}, j+\Delta j_{(k)}}\right)$. This produces a vector of normalized cross-correlations between the pattern and its $M$ neighbors, which we sort in descending order:

$$\mathbf{c} = \{c_{(m)}\}_{m=1}^{M} = \text{sort}(\{c_k\}_{k=1}^{M}), \quad c_{(1)} \geq c_{(2)} \geq \ldots \geq c_{(M)}.$$

Based on the above, we compute the moving difference in NCC scores ($c$):

$$\Delta c_{(m)} = \begin{cases} 0, & m=1 \\ c_{(m-1)} - c_{(m)}, & m=2,\ldots,M \end{cases}.$$

A plot of sorted NCC scores and the associated moving differences is shown in Figure S3. This preparation is done to identify the first statistically significant jump in NCC scores, thus defining a potential transition to a different PS variant. For each ordered neighbor index $m$, we define a two-sided local window excluding the current index:

$$\mathbf{\omega}_m = \{\Delta c_{(p)} : m - b_l \leq p < m\} \cup \{\Delta c_{(p)} : m < p \leq m + b_u\},$$

where $b_l$ and $b_u$ are, respectively, the lower and upper bounds of the search window (typically set to 8 and 5 for this work, respectively). The median of the $\Delta c$ values within this window is computed ($\tilde{\omega}_m$) along with the normalized median absolute deviation (as an estimate of the local deviation, which is resilient to outliers):

$$\sigma_m = 1.4826 \cdot \text{median}\left(|\mathbf{\omega}_m - \tilde{\omega}_m|\right).$$

A jump is identified if

$$\Delta c_{(m)} > \tilde{\omega}_m + z\sigma_m,$$

where $z$ is a threshold factor. The index of the first jump ($m^*$) is recorded. All patterns after the jump are not included in the neighbor pattern averaging, while all patterns before the first jump are included, given by the set: $\mathcal{K} = \{1, \ldots, m^* - 1\}$. Finally, the new averaged pattern is the weighted average of all neighboring patterns before the cutoff:

$$\hat{\mathbf{B}}_{i,j} = \frac{\mathbf{B}_{i,j} + \sum_{k \in \mathcal{K}} c_k \mathbf{B}_{i+\Delta i_{(k)}, j+\Delta j_{(k)}}}{1 + \sum_{k \in \mathcal{K}} c_k}.$$

This algorithm has been implemented in *python*, using *kikuchipy* to load the pattern file and *dask_mpi* to handle parallel processing of the large pattern arrays. Computations are performed in gridded blocks to retain the map navigation dimensions for neighborhood analysis while enabling memory-efficient parallel processing. This code will be made available upon publication of the manuscript.

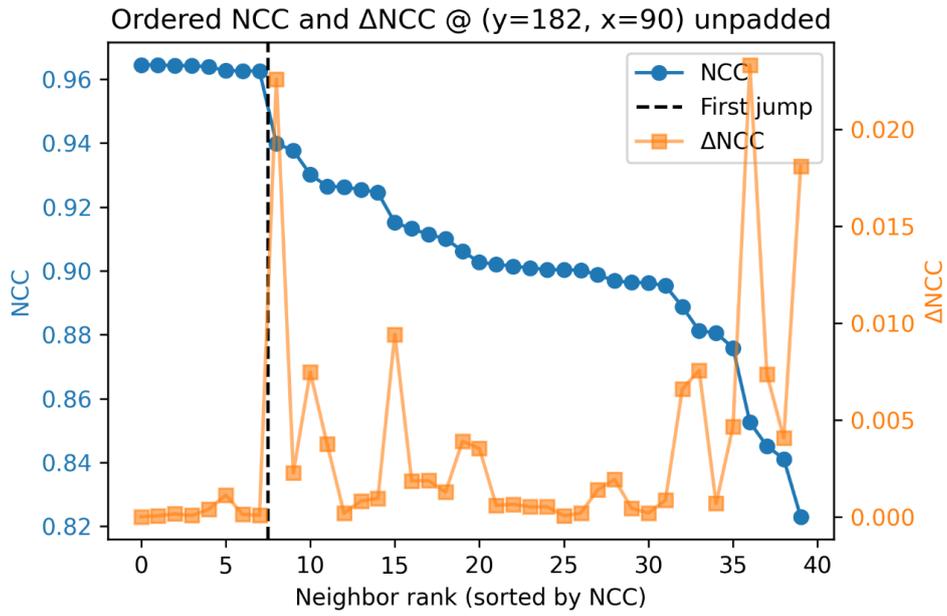

**Figure S3: NCC scores and moving differences for sorted neighboring patterns** (computed for pattern located at pixel y=182, x=90 in the BTO EBSD map, with a Manhattan search radius of 4).

## 5. Parameters for simulated patterns and reindexing examples

5.1 Parameters for simulated patterns

| Material | Lattice parameters [Å] | Space group | Atomic coordinates | Beam energy [kV] | Debye-Waller factor | Resolution [pixels] |
|---|---|---|---|---|---|---|
| BTO | $a$=3.99823 $c$=4.0262 | 99 | Ba: (0.0, 0.0, 0.0) Ti: (0.5, 0.5, 0.482) O1: (0.0, 0.5, 0.518) O2: (0.5, 0.5, 0.031) | 25 | 0.5 | 1001x1001 |
| PZT | $a$=4.055 $c$=4.1097 | 99 | Ti/Zr: (0.5, 0.5, 0.569) Pb: (0.0, 0.0, 0.0) O1: (0.5, 0.5, 0.0796) O2: (0.5, 0.0, 0.5996) | 15 | 0.5 | 1001x1001 |
| LiNbO$_3$ | $a$=5.14829 $c$=13.8631 | 161 | Nb: (0.0, 0.0, 0.0) O: (0.0509, 0.3587, 0.0527) Li: (0.0, 0.0, 0.225) | 20 | 0.5 | 1001x1001 |

**Table 5: Parameters for the simulation of spherical master patterns**

| Shape | Pixel size | binning | Detector tilt | azimuthal | Sample tilt | Pattern center |
|---|---|---|---|---|---|---|
| (338, 338) | 66.67 μm | 1x1 | 10° | -2° | 70° | (0.506, 0.194, 0.866) |

Table 6: Detector parameters for theoretical NCC landscapes

5.2 Pattern processing, PSS-NPA, and refined detector parameters for BTO scans

| Scan | DBS std | DBS trunc | FFT cutH | FFT cutL | AHE on | AHE clip | AHE nbins | AHE kernel | IQ (ΔIQ) | NCC (ΔNCC) |
|---|---|---|---|---|---|---|---|---|---|---|
| A1 | 23 | 2 | 3 | 50 | TRUE | 0.0001 | 128 | 80 | 0.947 (0.657) | 0.511 (0.436) |
| A2 | 20 | 2 | 3 | 50 | FALSE | - | - | - | 0.946 (0.67) | 0.512 (0.442) |
| A3 | 24 | 2 | 4 | 50 | FALSE | - | - | - | 0.947 (0.626) | 0.518 (0.431) |
| A4 | 15 | 3 | 3 | 50 | TRUE | 0.0001 | 512 | 64 | 0.946 (0.667) | 0.514 (0.443) |

Table 7: Optimized pattern processing parameters for the four EBSD scans on BTO

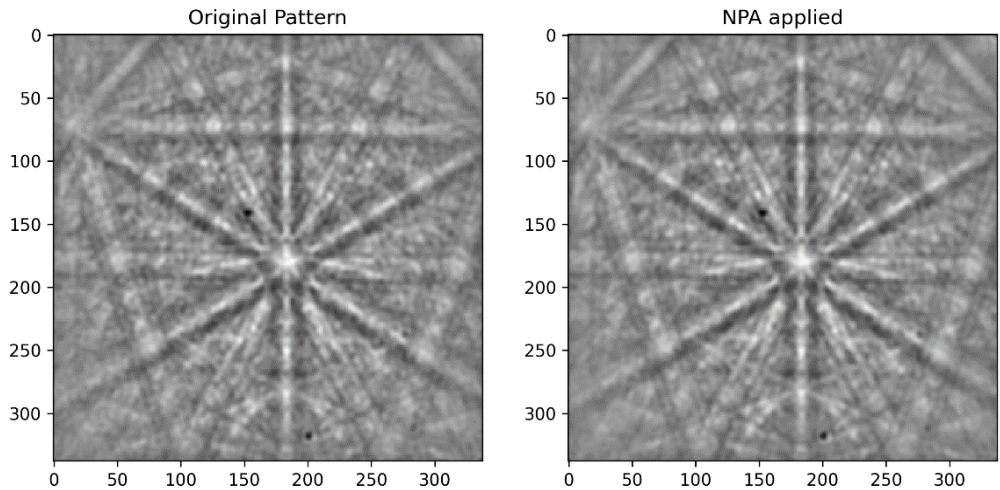

Figure S4: Example pattern before and after application of the PSS-NPA algorithm with a radius of 4 nearest neighbors.

| Shape | Pixel size | binning | Detector tilt | azimuthal | Sample tilt | Average pattern center |
|---|---|---|---|---|---|---|
| (338, 338) | 66.67 μm | 1x1 | 9.382° | -1.851° | 73.129° | (0.502, 0.125, 0.867) |

Table 8: Optimized detector for the BTO EBSD maps

5.3 Pattern processing, PSS-NPA, and refined detector parameters for PZT scans

| DBS std | DBS trunc | FFT cutH | FFT cutL | AHE on | AHE clip | AHE nbins | AHE kernel | IQ (ΔIQ) | NCC (ΔNCC) |
|---|---|---|---|---|---|---|---|---|---|
| 34 | 10 | 4 | 50 | TRUE | 0.005 | 512 | 128 | 0.938 (0.894) | 0.557 (0.256) |

Table 9: Optimized pattern processing parameters for the EBSD scans on PZT

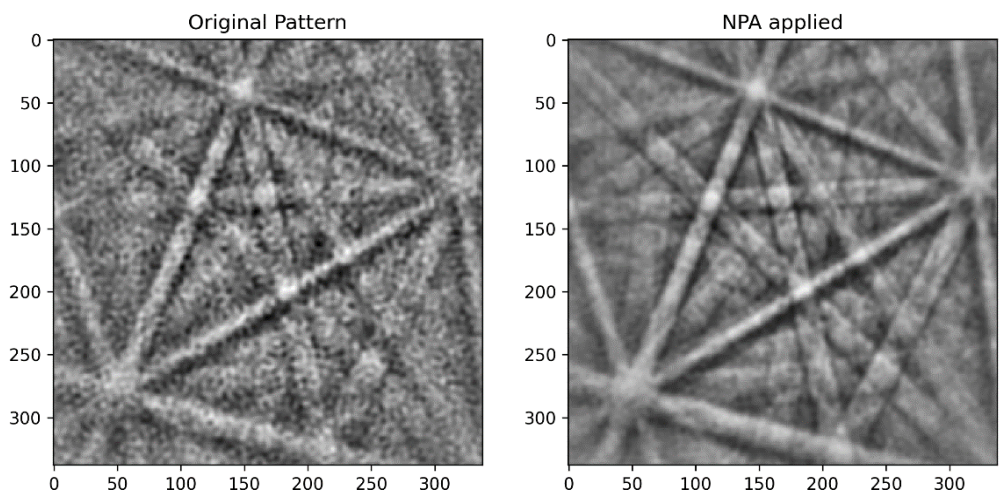

Figure S5: Example pattern before and after application of the PSS-NPA algorithm with a radius of 5 nearest neighbors.

| Shape | Pixel size | binning | Detector tilt | azimuthal | Sample tilt | Average pattern center |
|---|---|---|---|---|---|---|
| (338, 338) | 66.67 μm | 1x1 | 10° | -1.869° | 69.233° | (0.481, 0.234, 0.84) |

Table 10: Optimized detector for the PZT EBSD map (10x10 grid of patterns used for optimization)